\documentclass[aps, twocolumn, pre, floatfix, superscriptaddress,longbibliography]{revtex4-2}

\usepackage{amsmath,amssymb,amsthm}
\usepackage{bbold}
\usepackage{graphicx}

\renewcommand{\Im}{\operatorname{Im}}
\renewcommand{\Re}{\operatorname{Re}}

\newcommand{\tr}{\operatorname{tr}}

\graphicspath{{.}{figures/}}

\usepackage{color}

\definecolor{darkblue}{rgb}{0,0,.65}
\definecolor{darkgreen}{rgb}{0.3,0.6,0.3}
\definecolor{cyan1}{rgb}{0.0, 0.6, 0.6}
\usepackage[%
pdfstartview=FitH,%
breaklinks=true,%
bookmarks=true,%
colorlinks=true,%
anchorcolor=black,%
citecolor=red,
filecolor=black,%
menucolor=black,%
urlcolor=darkblue,%
linkcolor=blue,%
]{hyperref}

\usepackage{cleveref}

\newcommand{\ket}[1]{\ensuremath{\left|#1\right\rangle}}

\newcommand{\bc}{\boldsymbol{c}}
\newcommand{\bu}{\boldsymbol{u}}
\newcommand{\bphi}{\boldsymbol{\phi}}

\begin{document}

\title{The spectral boundary of the Asymmetric Simple Exclusion Process (ASEP) - free fermions, Bethe ansatz and random matrix theory}

\begin{abstract}
In non-equilibrium statistical mechanics, the Asymmetric Simple Exclusion Process (ASEP) serves as a paradigmatic example. 
We investigate the spectral characteristics of the ASEP, focusing on the spectral boundary of its generator matrix.  We examine finite ASEP chains of length $L$, under periodic (pbc) and open boundary conditions (obc). Notably, the spectral boundary exhibits $L$ spikes for pbc and $L+1$ spikes for obc.
Treating the ASEP generator as an interacting non-Hermitian fermionic model, we extend the model to have tunable interaction. In the non-interacting case, the analytically computed many-body spectrum shows a spectral boundary with prominent spikes.  For pbc, we use the coordinate Bethe ansatz to interpolate between the noninteracting case to the ASEP limit, and show that these spikes stem from clustering of Bethe roots.
The robustness of the spikes in the spectral boundary is demonstrated by linking the ASEP generator to random matrices with trace correlations or, equivalently, random graphs with distinct cycle structures, both displaying similar spiked spectral boundaries.

\end{abstract}

\newcommand{\dresdenTP}{Institut f\"{u}r Theoretische Physik, Technische Universit\"{a}t Dresden, D-01062 Dresden, Germany}
\newcommand{\maynooth}{Department of Theoretical Physics, Maynooth University, Co. Kildare, Ireland}

\author{Goran Nakerst}
\affiliation{\dresdenTP}
\affiliation{\maynooth}

\author{Toma\v{z} Prosen}
\affiliation{University of Ljubljana, Faculty for Mathematics and Physics, Jadranska 19, Ljubljana, Slovenia}

\author{Masudul Haque}
\affiliation{\dresdenTP}
\affiliation{\maynooth}
\affiliation{Max-Planck-Institut f\"{u}r Physik komplexer Systeme, D-01187 Dresden, Germany}

\maketitle

\section{Introduction}

The Asymmetric Simple Exclusion Process (ASEP) 
\cite{%
	Spitzer_AdvMath1970interacting_markov_process,Liggett_book1985interact_part_systems, Spohn_1991Dynamics_interacting_particles, 
	Schuetz_Domany_JourStatPhys1993asep, Derrida_Evans_Hakim_Pasquier_JPA1993, Derrida_PhysRep1998ASEP, Liggett_1999stochastic_interacting_systems, Schuetz_PhaseTransCritPhen2001, 
	Golinelli_Mallick_JourPhysA2006ASEP, Chou_Mallick_Zia_RepProgPhys2011, Mallick_PhysicaA2015_ASEP}  
is a well-studied paradigmatic stochastic many-body model that has been used to understand a wide range of non-equilibrium phenomena. This paper explores the spectral boundary of the markov matrix (the generator of ASEP), with a focus on a characteristic spiky formation, by establishing connections between the ASEP, non-interacting fermions, and random matrices featuring trace correlations.

The ASEP model has proven instrumental in shedding light on phenomena like non-equilibrium phase transitions
\cite{%
	Krug_PRL1991phase_transition, Derrida_Evans_Hakim_Pasquier_JPA1993, Sasamoto_JourPhysA_1999_asep_phases, Blythe_Essler_JourPhysA2000_asep_phases, Parmeggiani_Franosch_Frey_PRL2003_phase_coex, Blythe_Evans_JourPhysA2007mp_form},
and shock formation
\cite{%
    Derrida_et_al_JourStatPhys1993ASEP_shock, Mallick_JourPhysA1996_asep_impurity_shocks, 
    Jafarpour_PhysicaA2005_two_species_shocks, 
    Blythe_Evans_JourPhysA2007mp_form, 
    Kim_etal_JourStatMech2011_asep_shortcuts_shocks, 
    Arita_EPL2015_shocks},
among others.
Its versatility extends across various domains, such as protein synthesis
\cite{%
	MacDonald_Gibbs_Biopolymers1969, MacDonald_Gibbs_Pipkin_Biopolymers1968, Bressloff_Newby_RevModPhys2013intracellular_transport},
intracellular transport
\cite{%
	Bressloff_Newby_RevModPhys2013intracellular_transport, Neri_Kern_Parmeggiani_PRL2013_celltransport, Fang_et_al_RevModPhys2019_noneq_biology},
traffic flows
\cite{%
	Helbing_RevModPhys2001traffic},
and quantum dots
\cite{%
Karzig_vonOppen_PRB2010_quantum_dot}.
Another major incentive for its study is the association of the ASEP with interface dynamics and its connection to the Kardar-Parisi-Zhang equation in 1D (or equivalent noisy Burgers' equation) \cite{Meakin_et_al_PRA1986, Kardar_Parisi_Zang_PRL1986, Krug_Spohn_1991, Gwa_Spohn_PRA1992_Bethe_ansatz}.

The ASEP is a model where particles move stochastically on a one-dimensional lattice, adhering to exclusion interactions that restrict each site to a single particle, mirroring volume exclusion in real systems. Particles move to adjacent sites only if these sites are unoccupied. The process is termed 'asymmetric' due to the unequal probabilities for particle movement to the left or right, leading to directional bias. In cases where movement is limited to one direction, the model is referred to as Totally Asymmetric Simple Exclusion Process (TASEP).

A probability vector $P$ of particle configurations evolves according to the equation
\begin{equation}
    \frac{d}{dt} P(t) = H P(t),
\end{equation}
where $H$ is the generator matrix that governs the dynamics of the system. This markov (stochastic) matrix is a cornerstone of our study as it encapsulates all the dynamical information of the ASEP. The spectrum of $H$ is particularly insightful: it informs us about the various rates at which different states of the system evolve, which is crucial for understanding how the system approaches its steady state.

The asymmetry of the ASEP implies that the matrix $H$ is non-hermitian and its eigenvalues are generally complex. The real part of these eigenvalues relates to the relaxation times of eigenmodes, indicating how quickly the system returns to the steady state after a disturbance. The imaginary part, on the other hand, determines the oscillatory behavior of the system, setting the time scales of periodic or quasi-periodic patterns in the system evolution.

In this paper, we focus on finite chains of length $L$ and either periodic (pbc) or open boundary conditions (obc). The finite-dimensional nature of $H$ in these cases leads to a discrete and bounded spectrum. Analyzing this spectrum, especially establishing tight bounds on it, provides valuable insights into the aforementioned time scales and the overall dynamical properties of the system.

\begin{figure}
\begin{center}
	\includegraphics[width=\columnwidth]{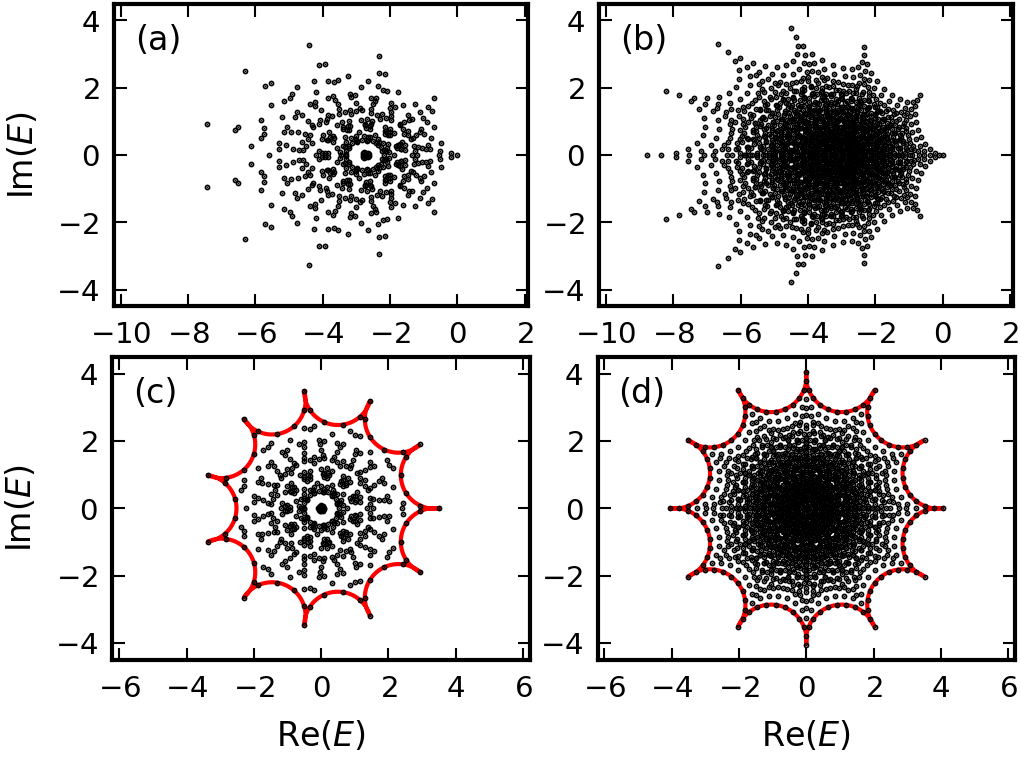}
	\caption{Spectrum of the generator matrix $H$ of TASEP \textbf{(a,b)} and the non-interacting TASEP \textbf{(c,d)} on $L=11$ sites. The spectrum shows $L$ spikes in \textbf{(a,c)} for pbc with $N=5$ particles and $L+1$ spikes in \textbf{(b,d)} for obc. Red solid lines in \textbf{(c,d)} denote the spectral boundary according to Eq.~\eqref{eq:H0_boundary}. \label{fig:spectrum_tasep}}
\end{center}
\end{figure}

Our primary objective is to investigate and explain an intriguing feature of the shape of the spectral boundary, namely, the prominent spikes clearly seen in Figs.\ \ref{fig:spectrum_tasep}(a,b) and also in previous studies \cite{Prolhac_JourPhysA2013_bulk_evs, Prolhac_JournStatMech2015_fluct_dev_TASEP}.  The  formation of these spikes -- $L$ spikes for pbc and $L+1$ for obc -- present a fascinating aspect of the spectral characteristics of the ASEP. Unraveling the mechanisms behind the formation of these spikes in the spectral boundary is a major focus of this work.
We elucidate the emergence of spectral spikes through three approaches. 

Firstly, the generator matrix $H$ is modeled as an interacting, non-Hermitian, spinless fermion system with interaction strength $U=1$. For $U=0$, $H$ reduces to a non-interacting fermion model.  Although this is not a Markov matrix, it is instructive to study the $U=0$ case as it is solvable as a non-Hermitian free-fermion Hamiltonian.  (We refer to this as the ``non-interacting ASEP''.)  The many-body spectrum of $H$ in this case, expressible as sums of single-particle eigenvalues on ellipses (circles for TASEP) in the complex plane, exhibits $L$ spikes ($L+1$ for obc) at its spectral boundary.  

In the second approach, we extend the coordinate Bethe ansatz method, traditionally used for calculating the spectrum of $U=1$ with pbc \cite{Gwa_Spohn_PRA1992_Bethe_ansatz}, to encompass arbitrary interaction strengths $U$. For TASEP, the many-body spectrum is constituted by sums of Bethe roots, which exhibit an elliptical clustering in the complex plane within the range $0\le U \le 1$. By focusing on the cluster sizes and disregarding finer Bethe root details, we demonstrate that the spectral boundary, akin to the $U=0$ case, is defined by sums of Bethe roots from neighboring clusters, resulting in a prominent display of $L$ spikes.

Lastly, we underscore the resilience of these spiky spectral boundaries by relating the TASEP to a random graph ensemble. In TASEP, the number of updates required to revert to a specific configuration is a multiple of $L$ ($L+1$ for obc) \cite{Prolhac_JourPhysA2013_bulk_evs}. We examine random graphs wherein all cycle lengths are divisible by $L$ ($L+1$ for obc). Our findings reveal that the spectral boundaries of both the adjacency matrix (analogous to $U=0$ in TASEP) and the Laplacian matrices (corresponding to $U=1$ in TASEP) of this random graph ensemble are characterized by the presence of $L$ ($L+1$) spikes.

The resilience of the spiky spectral boundary is noteworthy. This feature, inherent in the non-interacting fermion model, remarkably withstands the reintroduction of interactions. Furthermore, it prevails even when all aspects of $H$ are disregarded, except for the cycle lengths in the many-body graph.

The article is organized as follows: In Sec.~\ref{sec:generator} we introduce the generator matrix of ASEP with pbc and obc. In Sec.~\ref{sec:H0_pbc} and Sec.~\ref{sec:H0_obc} we present results of the non-interacting ASEP ($U=0$) with pbc and obc, respectively. In Sec.~\ref{sec:bethe} we investigate the interacting TASEP ($0\le U \le 1$) with pbc by Bethe ansatz. In Sec.~\ref{sec:random_graph} we compare TASEP to random graphs with the aforementioned cycle structure. We conclude in Sec.~\ref{sec:discussion}. Appendices \ref{sec:app_H0_obc} and \ref{sec:app_Mc} provide additional information on solving the non-interacting TASEP with obc. Appendix \ref{sec:app_bethe} details the derivation of Bethe equations for any $U$ with pbc, and Appendix \ref{sec:app_bethe_numerical} presents numerical specifics for solving these equations to determine the full spectrum of the generator matrix $H$.

%%%%%%%%%%%%%%%%%%%%%%%%%%%%%%%%%%%%%%%%%%%%%%%%%%%%%%%%%%%%%%%%%%%%%%%%%%%%%%%%%%%%%
%%%%%%%%%%%%%%%%%%%%%%%%%%%%%%%%%%%%%%%%%%%%%%%%%%%%%%%%%%%%%%%%%%%%%%%%%%%%%%%%%%%%%
%%%%%%%%%%%%%%%%%%%%%%%%%%%%%%%%%%%%%%%%%%%%%%%%%%%%%%%%%%%%%%%%%%%%%%%%%%%%%%%%%%%%%
\section{Generator matrix of ASEP}\label{sec:generator}

In this section, we will introduce the generator matrix $H$ of the ASEP for pbc and obc as non-Hermitian fermion models, along with essential notation.

We consider ASEP chains of length $L$. The number of particles in the chain is denoted by $N$ and the particle density by $\rho=N/L$. The probability for a particle to hop right or left in time $dt$ is $p\,dt$ or $q\,dt$,
respectively, with the convention $p+q=1$ unless specified otherwise.

Let us introduce
\begin{equation}
    H = H_U = H_0 + U \mathcal{I},
\end{equation}
where $H_0$ is a matrix with non-negative off-diagonal elements and zero diagonal entries and $\mathcal{I}$ is a diagonal matrix. The term $U$ denotes the interaction strength. In the ASEP context, $H_0$ represents a non-interacting spinless fermion model, and $\mathcal{I}$ a 4-point (2-body) fermion interaction. 

The generator of the ASEP is $H=H_1$ with interaction strength $U=1$. Here, $H_1$ is the generator of a stochastic Markov process and a stochastic matrix, where the sums of all columns of $H_1$ equal zero. This property is ensured by the diagonal elements of $\mathcal{I}$ equalling the sums of the corresponding columns of $H_0$,
\begin{equation}\label{eq:markov_property}
    \mathcal{I}_{jj} = \sum_{k} \left(H_0\right)_{kj}.
\end{equation}
Whenever $U\neq1$, $H_U$ ceases to be a Markov matrix and does not generate the ASEP or any other stochastic process.

Studying $H_U$ with $U\neq 1$ could elucidate the $U=1$ case for two reasons.
First, the analyticity of $H_U$ in $U$ suggests that its properties at $U\neq 1$ could be extrapolated to $U=1$. 
Second, the diagonal matrix $\mathcal{I}$ exists only to ensure the Markov property of $H_1$ and, according to Eq.~\eqref{eq:markov_property}, is entirely determined by $H_0$. Therefore, ignoring $\mathcal{I}$ in the $U=0$ case likely retains some features of the Markov matrix $H_1$.

\subsection{Periodic Boundary Conditions}

For pbc the matrices $H_0$ and $\mathcal{I}$ are given by
\begin{align}
	H_0 &= \sum_{j=1}^L \left(p\sigma_{j+1}^+ \sigma_j^- + q\sigma_j^+ \sigma_{j+1}^- \right) \label{eq:H0_pbc_def}\\
	\mathcal{I} &= \frac{1}{4}\sum_{j=1}^L \left( \sigma_j^z \sigma_{j+1}^z - 1\right). \label{eq:I_pbc_def}
\end{align}
The symbols $\sigma^\pm$ denote spin raising and lowering operators, while $\sigma^z$ denotes the $z$-component of the spin. The spin-up state is interpreted as a particle present, while the spin-down state is interpreted as a particle absent.

Without loss of generality we can assume $q\le p$. For $p,q\neq 0$ the matrix $H$ can be mapped to an XXZ spin $1/2$ chain with non-Hermitian, twisted boundary conditions \cite{Golinelli_Mallick_JourPhysA2006ASEP}. For $p=q$ the matrix $H$ is Hermitian and for $U=1$ reduces to the Heisenberg spin chain. 

The matrix $H$ can be written in terms of fermions by a Jordan-Wigner transformation
\begin{equation}
    c_j^{(\dagger)}=e^{i\pi \sum_{k<j} \sigma_k^+\sigma_k^-} \sigma_j^{-(+)},
\end{equation}
where $c^{(\dagger)}_j$ are fermionic annihiliation (creation) operators. The corresponding fermionic operator $H$ is then given by
\begin{align}
	H_0 &=\sum_{j=1}^{L-1} \left( p c_{j+1}^\dagger c_j + q c_j^\dagger c_{j+1} \right)
    + (-1)^{N+1} (p c_1^\dagger c_L + q c_L^\dagger c_1) \label{eq:H0_pbc_c} \\
	\mathcal{I} &= \sum_{j=1}^L c_j^\dagger c_j c_{j+1}^\dagger c_{j+1} - N. \label{eq:I_pbc_c}
\end{align}
$H_0$ is the Hamiltoninian of non-hermitian free spinless fermions, while $\mathcal{I}$ denotes a fermionic quartic interaction. The off-diagonal elements of $H$ given by $H_0$ are non-negative, while the diagonal of the diagonal matrix $\mathcal{I}$ consists of non-positive values. 

The number of particles $N$ (spin-up states) is conserved by $H$ for all interaction strengts $U$.

\subsection{Open Boundary Conditions}
For obc, the matrix $H_0$ is given by
\begin{equation}\label{eq:asep_obc_def}
	H_0 = \sum_{j=1}^{L-1} \left(p\sigma_{j+1}^+ \sigma_j^- + q\sigma_j^+ \sigma_{j+1}^- \right) + \alpha \sigma_1^+ + \gamma \sigma_1^- + \beta \sigma_L^- + \delta \sigma_L^+,
\end{equation}
while the diagonal $\mathcal{I}$ is given by
\begin{align}
    \mathcal{I}
    &= \frac{1}{4}\sum_{j=1}^{L-1} \left( \sigma_j^z \sigma_{j+1}^z - 1\right) \notag\\
    &+ \frac{1}{2}\left[ (p-q-\alpha+\gamma)\sigma^z_1 + (q-p-\delta+\beta)\sigma^z_L \right] \notag \\
    &-\frac{1}{2}\left[ \alpha+\beta+\gamma+\delta \right].
\end{align}
The bulk term of $H_0$ for obc is the same as for pbc. The terms at the edges of the chain on site 1 and $L$ with parameters $\alpha,\beta,\gamma,\delta$ denote particles hopping in and out of the chain from an infinite reservoir of particles. Similar to pbc and $p,q\neq 0$, $H$ can be mapped to an XXZ chain with non-Hermitian, twisted boundary conditions \cite{Essler_Rittenberg_JournPhysA_1996_asep_xxz_obc}.

As in the pbc case, the operator $H$ can be written in terms of fermions. The single spin operators at the end of the chain on site 1 and $L$ hinder a straightforward application of a Jordan-Wigner transformation. Instead, we treat the infinite reservoir as an additional site. We enlarge the chain of length $L$ to a ring of length $L+1$ and change the terms connecting to site $L+1$ accordingly. This is formally done by application of the well-known Kramers-Wannier duality transformation \cite{Kogut_RevModPhys_1979} $\sigma_j^x \to \prod_{l=1}^j \sigma_l^z$ and $\sigma_j^z \to \sigma_j^x \sigma_{j+1}^x$. The details are in Appendix \ref{sec:app_H0_obc}. Adding a site to the chain comes with the caveat that the multiplicity of every eigenvalue of the so-transformed $H_0$ is doubled.

To keep the algebra simpler we restrict to the TASEP case $p=1$ and $q=\gamma=\delta=0$, leaving $\alpha$ and $\beta$ as free parameters. The following results can be straightforwardly generalized to arbitrary $p,q,\gamma,\delta$. As outlined in Appendix~\ref{sec:app_H0_obc} the Hamiltonian $H_0$ is expressible in terms of spinless fermions $c,c^\dagger$ as
\begin{align}
    H_0 = &\alpha(c_{L+1} - c_{L+1}^\dagger) c_1^\dagger 
	+ \sum_{j=1}^{L-1} \left[ c_j c_{j+1}^\dagger \right] \nonumber \\
	&+ (-1)^{L} \mathcal{P}_c \beta c_L(c_{L+1}+c_{L+1}^\dagger),\label{eq:H0_obc_c}
\end{align}
where $\mathcal{P}_c$ denotes the parity of the fermion number 
\begin{align}
\mathcal{P}_c = (-1)^{\sum_{j=1}^{L+1}c^{\dagger}_j c_j} = (-1)^N,
\end{align}
which is conserved by $H_0$. Restricted to a fixed parity sector, $H_0$ is a quadratic Hamiltonian. The corresponding spectrum is the same for each parity sector leading to the aforementioned doubling of the spectral multiplicity. This will be shown in detail in Sec.~\ref{sec:H0_obc_spectrum}.

In summary, the non-interacting TASEP $H_0$ on $L$ sites with obc can be written as a free fermion model on $L+1$ sites, with twisted pbc and `superconducting' terms $c_{L,1}^{(\dagger)} c_{L+1}^{(\dagger)}$ connecting to the additional site $L+1$.

\subsection{Spectrum}

All eigenvalues of $H$ are either real or come in complex conjugate pairs. This characteristic stems from the fact that $H$ can be represented as a real matrix. Specifically, for the case where $U=1$, the stochastic nature of $H$ dictates that its spectrum is situated in the left half of the complex plane.

Fig.~\ref{fig:spectrum_tasep} presents the spectrum of TASEP on a lattice with $L=11$ sites. The spectral boundary shows $L$ spikes for pbc ($N=5$ particles) for $U=1$ in (a) and $U=0$ in (c) and $L+1$ spikes for obc and $U=1$ in (b) and $U=0$ in (d). For obc the parameters corresponding to the reservoirs are chosen as $\alpha=\beta=1$ and $\gamma=\delta=0$. The subsequent sections primarily aim to derive the mechanism responsible for the spikes in the spectral boundary.

Panels (c) and (d) of Fig.~\ref{fig:spectrum_tasep} reveal a highly structured spectrum for the non-interacting TASEP $H_0$, exhibiting rotational invariance at angles $2\pi/L$ for pbc and $2\pi/(L+1)$ for obc. This characteristic stems from a ``quasi-symmetry'' of $H_0$, which is investigated in detail in Sections~\ref{sec:H0_pbc} and~\ref{sec:H0_obc}.

For TASEP with obc, the spectral boundary spikes are always prominent, as illustrated for the non-interacting TASEP in Sec.~\ref{sec:H0_obc}. However, this is not the case for pbc. In Fig.~\ref{fig:spectrum_tasep_small_N}(a) the spectrum of the pbc TASEP ($U=1$) and in (b) its non-interacting variant ($U=0$) are presented for $L=40$ sites and $N=2$ particles, without any noticeable spikes in the spectral boundary. Sec.~\ref{sec:H0_pbc} will demonstrate that, technically, the spectral boundary of the non-interacting TASEP has $L=40$ spikes, but their distinctiveness fades in the dilute limit where $\rho\to0$.

%%%%%%%%%%%%%%%%%%%%%%%%%%%%%%%%%%%%%%%%%%%%%%%%%%%%%%%%%%%%%%%%%%%%%%%%%%%%%%%%%%%%%
%%%%%%%%%%%%%%%%%%%%%%%%%%%%%%%%%%%%%%%%%%%%%%%%%%%%%%%%%%%%%%%%%%%%%%%%%%%%%%%%%%%%%
%%%%%%%%%%%%%%%%%%%%%%%%%%%%%%%%%%%%%%%%%%%%%%%%%%%%%%%%%%%%%%%%%%%%%%%%%%%%%%%%%%%%%
\section{``Non-interacting'' ASEP with pbc}\label{sec:H0_pbc}

In this section, we investigate the spectrum of the non-interacting ASEP $H_0$ for pbc given by Eq.~\eqref{eq:H0_pbc_def} and Eq.~\eqref{eq:H0_pbc_c}, respectively. Sec.~\ref{sec:H0_pbc_sb} is devoted to the calculation of the single-body eigenvalues of $H_0$. In Sec.~\ref{sec:H0_pbc_rot} we show the rotational invariance of the many-body spectrum of TASEP and in Sec.~\ref{sec:H0_pbc_boundary} we combine the results from the preceding subsections and show how the spiky spectral boundary emerges. We quantify the prominence of the spikes in Sec.~\ref{sec:H0_pbc_spikes} and comment on whether they survive in the limit of large $L$. 

\subsection{Single-body spectrum}\label{sec:H0_pbc_sb}

\begin{figure}
\begin{center}
	\includegraphics[width=\columnwidth]{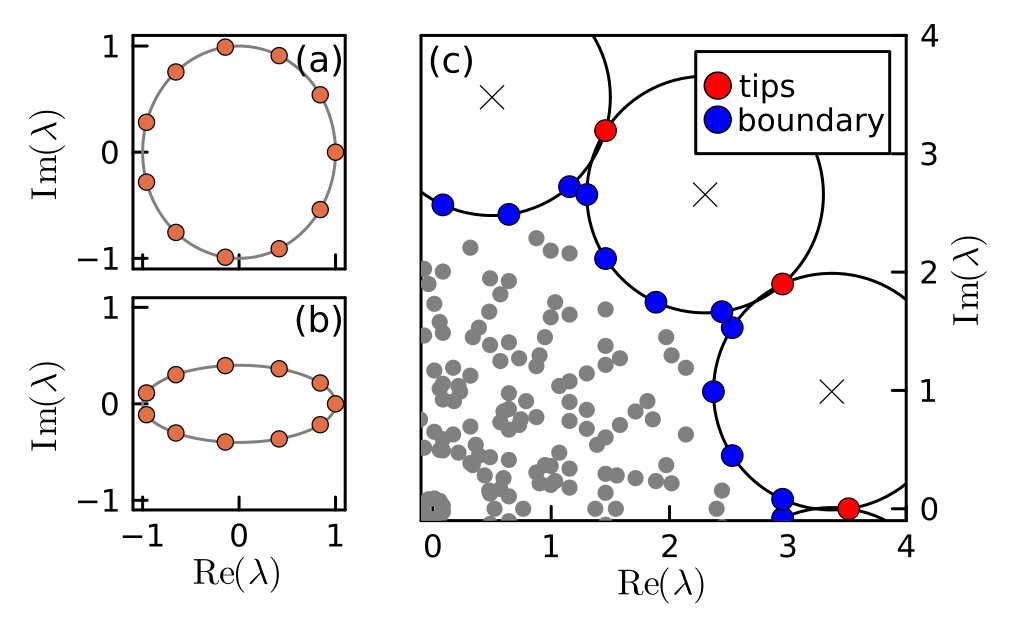}
	\caption{Spectrum of the non-interacting TASEP $H_0$ on $L=11$ sites with pbc. Single-body eigenvalues with $p=1$ and $q=0$ in \textbf{(a)} and $p=0.7$ and $q=0.3$ in \textbf{(b)}.  In \textbf{(c)} we show part of the many-body spectrum with $N=5$ particles highlighting the tips of the spikes (red) and other boundary eigenvalues (blue).  All boundary eigenvalues are located on circles of radius $1$, with crosses marking the midpoints.
 \label{fig:H0_pbc}}
\end{center}
\end{figure}

Let us focus on the totally asymmetric case $p=1$ and $q=0$ first.  Considering the single-body sector of $H_0$ as given in Eq.~\eqref{eq:H0_pbc_c},  we see that the single-body spectrum $\lambda$ is given by roots of the polynomial
\begin{equation}
    \lambda^L + (-1)^{N+1}.
\end{equation}
The roots are given by $\lambda=\omega^j$, where $\omega= e^{i\pi/L}$ and $0\le j<2L$ runs over all even (odd) integers when $N$ is odd (even). Thus the single-body spectrum lies on the unit circle. In Fig.~\ref{fig:H0_pbc}(a) the single-body spectrum for $p=1$ and $q=0$ and $L=11$ and odd $N$ is shown together with the unit circle.

For arbirtrary values of $p$ and $q$, the single-body spectrum is represented as
\begin{equation}\label{eq:H0_pbc_sb}
    \lambda = p \omega^j + q \omega^{-j},
\end{equation}
with $j$ defined as previously. This spectrum lies on an ellipse with foci at $\pm 2\sqrt{pq}$ and semi-major axis $p+q$ and semi-minor axis $p-q$,
\begin{equation}\label{eq:H0_ellipse}
    \{(p+q)\cos(t) + i (p-q) \sin(t) : 0\le t\le 2\pi \}.
\end{equation}
Figure~\ref{fig:H0_pbc} (b) illustrates the single-body spectrum for $p=0.7$ and $q=0.3$, alongside the ellipse defined by Eq.~\eqref{eq:H0_ellipse}.

The structure of the single-body spectrum for any $p, q$ suggests a straightforward relation with the totally asymmetric scenario $q=0$. By modifying the imaginary component while maintaining the real part constant,
\begin{equation}
z \to \Re z + i \frac{p+q}{p-q}\Im z,
\end{equation}
we can convert the single-body eigenvalues for general $p, q$ values to those corresponding to the $q=0$ case. This transforms the ellipse into a circle of radius $p+q$. Without loss of generality, we restrict ourselves to $p=1$ and $q=0$ for the remainder of this section.

\subsection{Rotational invariance}\label{sec:H0_pbc_rot}
With $p=1$ and $q=0$, the single-body spectrum remains unchanged under complex plane rotations of $2\pi/L$. This rotational invariance also applies to the many-body spectrum, which comprises sums of single-body eigenvalues.

Furthermore, this symmetry is evident in $H_0$ when transforming spin and fermionic operators. Transforming $c_j\to e^{-i2\pi j/L} c_j =\tilde{c}_j$ and $c_j^\dagger \to e^{i2\pi j/L} c_j^\dagger = \tilde{c}_j^\dagger$, or in terms of spin operators $\sigma_j^{\pm} \to e^{\pm i2\pi j/L} \sigma_j^{\pm} = \tilde{\sigma}_j^{\pm}$, results in $e^{i2\pi/L} H_0 = \tilde{H}_0$. Here, $\tilde{H}_0$ is constructed like $H_0$, but using the modified operators $\tilde{c}, \tilde{c}^\dagger$ ($\tilde{\sigma}^\pm$). Since these altered operators maintain their respective (anti-)commutation relations, the spectra of $H_0$ and $\tilde{H}_0$ are identical. Therefore, the spectrum of $H_0$ is invariant under $2\pi/L$ rotations.

\subsection{Spectral boundary} \label{sec:H0_pbc_boundary}

The structure of the many-body spectrum as observed in Fig.~\ref{fig:spectrum_tasep} is now a consequence of the relation of single-body to many-body eigenvalues and the rotational symmetry. 

For ease of notation, we define $\lambda_j = \omega^{2j}$ when $N$ is odd, and $\lambda_j = \omega^{2j+1}$ for even $N$. The many-body eigenvalues are obtained by adding $N$ of these $L$ single-body eigenvalues.  More precisely, the  many-body eigenvalues $E$ correspond uniquely to configurations $s=(s_1,\dots,s_L)\in\{0,1\}^L$, where $\sum_j s_j = N$, and are given by
\begin{equation}
    E = \sum_{j=1}^L s_j \lambda_j.
\end{equation}
The many-body eigenvalues $E_t$ which appear at the spike tips, have the highest absolute values and are derived from configurations $s$ with contiguous non-zero $s_j$ entries. Specifically, each of the $L$ tips $E_t(j_0)$ is linked to an index $1\le j_0 \le L$ and a configuration $s=s_t(j_0)$ with
\begin{equation}
    s_j = \begin{cases}
        1 & j_0 \le j \le j_0+N-1 \\
        0 & \text{otherwise.}
    \end{cases}
\end{equation}
Here, $j \equiv j - L$ is applied for $j>L$. The eigenvalues $E_t(j_0)$ are calculated as
\begin{equation}\label{eq:H0_pbc_Et}
    E_t(j_0) = \sum_{j=j_0}^{j_0+N-1} \lambda_j.
\end{equation}
Configurations $s$ that lead to spike tips are termed 'domain wall configurations'. The many-body eigenvalues $E_t$ are depicted as red circles (light colored in print) in Fig.~\ref{fig:H0_pbc}(c).

Boundary eigenvalues in the many-body spectrum arise from 'interpolating' between configurations of adjacent spike tips. In these configurations, the domain walls differ by a shift of one site. The interpolation process between these two domain walls involves moving a single particle (or executing a single spin flip). As a result, the configurations formed contain a maximum of two domain walls, each separated by one site. 
Specifically, boundary configurations $s=s_b(j_0,l_0)$ are associated with indices $1\le j_0\le L $ and $j_0\le l_0 \le j_0+N$, defined as
\begin{equation}
    s_j = \begin{cases}
        1 & j_0 \le j \le j_0+N \text{ and } j\neq l_0 \\
        0 & j=l_0 \\
        0 & \text{otherwise.}
    \end{cases}
\end{equation}
Again, $j \equiv j - L$ is used for $j>L$. The corresponding boundary eigenvalues $E_b(j_0, l_0)$ are computed by
\begin{equation}\label{eq:H0_pbc_Eb}
    E_b(j_0, l) = \sum_{j=j_0; j\neq l}^{j_0+N} \lambda_j.
\end{equation}
When $l_0=j_0$ or $l_0=j_0+N$ (indicating a single domain wall), the boundary eigenvalue matches a spike tip, $E_b(j_0,j_0)=E_t(j_0+1)$ or $E_b(j_0, j_0+N)=E_t(j_0)$, respectively. The boundary eigenvalues $E_b(j_0,l)$ for $j_0<l<j_0+N$ are those many-body eigenvalues located 'between' the spike tips $E_t(j_0)$ and $E_t(j_0+1)$, depicted as blue circles in Fig.~\ref{fig:H0_pbc}(c).

Eq.\eqref{eq:H0_pbc_Eb} can be reformulated as
\begin{equation}
E_b(j_0, l) = \sum_{j=j_0}^{j_0+N} \lambda_j - \lambda_l.
\end{equation}
Given $|\lambda_l|=1$ and the independence of the sum from $l$, all boundary eigenvalues are on $L$ circles of radius 1. For $N\le L/2$, the circle midpoints are the many-body spectrum tips $E_t^{(N+1)}(j_0)$ with $N+1$ particles. The tips $E_t^{(N)}$ intersect two adjacent circles. This is illustrated in Fig.~\ref{fig:H0_pbc}(c) with circles as black lines and midpoints as gray crosses.

According to Eq.~\eqref{eq:H0_pbc_Et}, all tips reside on a circle with radius $R$, defined as
\begin{equation}\label{eq:H0_pbc_R}
    R = \left|\frac{1-e^{i2\pi N/L}}{1-e^{i2\pi/L}}\right| = \frac{\sin(\pi N/L)}{\sin(\pi/L)}.
\end{equation}
This radius, combined with the circular pattern of the boundary eigenvalues, enables us to establish a continuous boundary for the many-body spectrum. It is formed by the intersection of all circles of radius 1 with the disc of radius $R$ from Eq.~\eqref{eq:H0_pbc_R}. The boundary is parameterized by
\begin{equation}\label{eq:H0_boundary}
	z_B(t) = e^{-if(t)}\left(\gamma_1 + \gamma_2 e^{ig(kt)} \right),
\end{equation}
with $\gamma_1 = \frac{\sin(\pi\rho)}{\sin(\pi/L)}$ and $\gamma_2 = 1$, with piece-wise constant $f$,
\begin{equation}
	f(t) = \frac{\pi}{L} \left( 2 \left\lfloor \frac{Lt}{2\pi} \right\rfloor - 1\right),
\end{equation}
and $g$ is piece-wise the identity,
\begin{equation}
    g(t) = \pi(1-\rho) + \rho (t \operatorname{mod}\ 2\pi).
\end{equation}
The continuous boundary $z_B(t)$ is illustrated as a red (gray in print) curve in Fig.~\ref{fig:spectrum_tasep}(c) for $L=11$ and $N=5$ and in Fig.~\ref{fig:spectrum_tasep_small_N}(b) for $L=40$ and $N=2$. As expected, all boundary eigenvalues reside on the continuous boundary parametrized by $z_B(t)$.

Eq.~\eqref{eq:H0_boundary} is related to the spectral boundary of random matrices with higher-order cyclic correlations between $L$-tuples of matrix elements, akin to random graphs with a dominant cycle structure \cite{Aceituno_Rogers_Schomerus_PRE2019}. Their spectral boundary forms a hypotrochoidic curve, which is recovered from Eq.~\eqref{eq:H0_boundary} by letting $f(t)=g(t)=t$. This relation hints at the connection between the spectral boundary of the non-interacting TASEP and random matrices; we explore this connection in Section \ref{sec:random_graph}.

\subsection{Quantification of spikes}\label{sec:H0_pbc_spikes}

This subsection aims to measure the sharpness of the spectral boundary in the non-interacting TASEP, particularly focusing on whether spikes persist in large system sizes and, if so, how. For simplicity, we consider particle densities $0\le \rho \le 1/2$. As the ASEP spectrum is invariant under changing $\rho\to 1-\rho$ this comes with no loss of generality.

To assess the spikiness of the spectral boundary, we examine the ratio between two distances: $d_t$, the distance between spike tips, and $d_b$, the maximum extension of the spectral boundary beyond a circle of radius $R$. Recall, this circle of radius $R$ represents the smallest enclosing disk for the TASEP spectrum. $d_b$ measures how far the radius 1 circles, carrying the boundary eigenvalues, reach into the enclosing circle. A larger $d_b$ relative to $d_t$ indicates that these radius 1 circles extend more into the enclosing spectrum. Therefore, the ratio $2d_b/d_t$ quantifies the spikiness of the boundary. A value close to 1 suggests a spiky boundary, while a significantly smaller ratio implies a less spiky boundary. This factor of two arises because $d_t$ pertains to the diameter of the boundary circles, whereas $d_b$ is compared to their radius.

Following some simple trigonometry one finds that the distances $d_t$ and $d_b$ are given by
\begin{equation}\label{eq:H0_pbc_dt}
    d_t = 2\sin(\pi \rho )
\end{equation}
and
\begin{equation}\label{eq:H0_pbc_db}
    d_b = 1 - \frac{\cos(\pi\rho + \pi/(2L))}{\cos(\pi/(2L))}.
\end{equation}
The fraction $2d_b/d_t$ then simplifies to
\begin{equation}\label{eq:H0_pbc_db_dt}
    \frac{2d_b}{d_t} = \tan(\pi\rho/2) + \tan(\pi/(2L)).
\end{equation}
Eq.~\eqref{eq:H0_pbc_db_dt} shows a monotonic increase with $\rho$, indicating that the spectral boundary becomes more pronouncedly spiky at higher $\rho$ values. Due to the invariance of the spectrum under the transformation $\rho \to 1-\rho$, the boundary reaches its maximum spikiness at $\rho=1/2$.

The analytical findings are confirmed by panels (c) and (b) in Fig.~\ref{fig:spectrum_tasep} and Fig.~\ref{fig:spectrum_tasep_small_N}, respectively. In Fig.~\ref{fig:spectrum_tasep}(c), the many-body spectrum of $H_0$ is markedly spiky for $\rho=5/11\approx 0.45$, whereas in Fig.~\ref{fig:spectrum_tasep_small_N}(b), the spectral boundary is nearly circular, aligning with the low $\rho$ value of 2/40=0.05.

In examining the large $L$ limit, we will explore two scenarios: the ``thermodynamic'' limit, where both $N$ and $L$ increase to infinity while maintaining a fixed $\rho$, and the few-particle (dilute) limit, where $N$ remains constant and only $L$ approaches infinity.

\subsubsection{``Thermodynamic'' limit}
In the thermodynamic limit, the distance $d_t$ remains constant, whereas $d_b$ approaches $1-\cos(\pi/\rho)$. Consequently, the ratio $2d_b/d_t$ tends towards $\tan(\pi\rho/2)$. This implies that for any non-zero $\rho$, the spiky structure of the spectral boundary is preserved in the thermodynamic limit, becoming more pronounced with increasing $\rho$.

Fig.~\ref{fig:spectrum_tasep}(c) presents the many-body spectrum of the non-interacting TASEP for $L=11$ and $N=5$, with Fig.~\ref{fig:H0_pbc}(c) offering a closer view of the spectral boundary. Here, $\rho\approx0.45$ and $2d_b/d_t\approx 1.01$ indicate pronounced spikes of the spectral boundary, as evident.

Regarding the length scales at which these spikes are observable, consider the following: The radius $R$ of the spectrum scales as $O(L)$, necessitating a rescaling of the spectrum by $1/L$ to ensure a well-defined spectral density in the thermodynamic limit. At an infinite $L$, this rescaled spectrum densely fills the unit circle. For finite $L$, the tips of the spikes are spaced at a distance of $d_t=O(1/L)$, and the distance $d_b$ of the spectral boundary from the unit circle is also $O(1/L)$. Therefore, at the length scale of $1/L$, the spiky nature of the spectral boundary is distinctly visible.

\subsubsection{Dilute limit (large $L$, constant $N$)}

\begin{figure}
\begin{center}
	\includegraphics[width=\columnwidth]{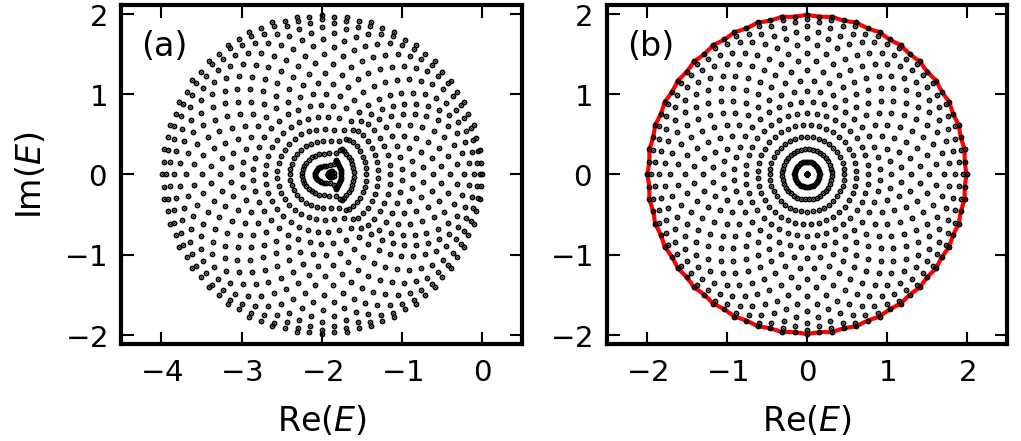}
	\caption{Spectrum of the generator matrix $H$ of TASEP \textbf{(a)} and the ``non-interacting'' TASEP \textbf{(b)} on $L=40$ sites with $N=2$ particles (dilute limit). The red solid line in \textbf{(b)} denotes the spectral boundary according to Eq.~\eqref{eq:H0_boundary}. The spectral boundary appears smooth and non-spiky in both panels.\label{fig:spectrum_tasep_small_N}}
\end{center}
\end{figure}

In the scenario where $N$ is fixed and $L$ increases, both distances $d_t$ and $d_b$ decrease, scaling as $O(1/L)$ and $O(1/L^2)$, respectively. Consequently, the ratio $2d_b/d_t$ tends towards 0, as indicated by Eq.~\eqref{eq:H0_pbc_db_dt}. Therefore, in this limit, the spiky structure of the spectral boundary does not persist. 

Fig.~\ref{fig:spectrum_tasep_small_N} shows the many-body spectrum of the TASEP for $L=40$ and $N=2$, representative of the dilute limit.  We show both a TASEP case ($U=1$) and a non-interacting TASEP case ($U=0$).   With a $2d_b/d_t$ ratio of $\approx 0.01$ it reveals a non-spiky spectral boundary, barely distinguishable from a circle, as shown by the red curve (gray in print) in Fig.~\ref{fig:spectrum_tasep_small_N}(b).

%%%%%%%%%%%%%%%%%%%%%%%%%%%%%%%%%%%%%%%%%%%%%%%%%%%%%%%%%%%%%%%%%%%%%%%%%%%%%%%%%%%%%
%%%%%%%%%%%%%%%%%%%%%%%%%%%%%%%%%%%%%%%%%%%%%%%%%%%%%%%%%%%%%%%%%%%%%%%%%%%%%%%%%%%%%
%%%%%%%%%%%%%%%%%%%%%%%%%%%%%%%%%%%%%%%%%%%%%%%%%%%%%%%%%%%%%%%%%%%%%%%%%%%%%%%%%%%%%
\section{``Non-interacting'' TASEP with obc}\label{sec:H0_obc}
In this section, we will present the analytical derivation of the many-body spectrum of the non-interacting TASEP $H_0$ with obc, specifically for $p=1$ and $q=\gamma=\delta=0$. Generalizations to arbitrary $p,q,\gamma,\delta$ are straightforward. 

In Sec.~\ref{sec:H0_obc_rot} we establish the rotational invariance of the spectrum of $H_0$. In Sec.~\ref{sec:H0_obc_spectrum} we derive its single-particle spectrum and demonstrate its relation to the many-body eigenvalues. Sec.~\ref{sec:H0_obc_boundary} demonstrates that the spectral boundary of $H_0$, similar to the pbc case, is defined by the intersection of circles with a disk, featuring $L+1$ spikes. In the limit of large $L$, this boundary is akin to the pbc case with density $\rho=1/2$, highlighted in Sec.~\ref{sec:H0_obc_spikes}.

\subsection{Rotational symmetry}\label{sec:H0_obc_rot}
The spectrum of the non-interacting TASEP $H_0$ is invariant under rotations of angle $\frac{2\pi}{L+1}$. Similar to the pbc case, consider the change of operators $c_j^\dagger \to e^{i\frac{2\pi}{L+1}j} c_j^\dagger = \tilde{c}_j^\dagger$ and $c_j \to e^{-i\frac{2\pi}{L+1}j} c_j = \tilde{c}_j$ or, equivalently, $\sigma_j^\pm \to e^{\pm i\frac{2\pi}{L+1} j}\sigma_j^\pm = \tilde{\sigma}^\pm$. This change implies that $e^{i\frac{2\pi}{L+1}} H_0 = \tilde{H}_0$, where $\tilde{H}_0$ is $H_0$ with $c,c^\dagger$ ($\sigma$) replaced by the tilde operators. As the tilde operators fulfill the canonical (anti-)commutation relations of fermion operators (Pauli matrices), the spectrum of the non-interacting TASEP is invariant under rotations of angle $\frac{2\pi}{L+1}$.

\subsection{Single- and many-body spectrum}\label{sec:H0_obc_spectrum}

Before we diagonalize $H_0$, let us specify the parity sector as $s=(-1)^L \mathcal{P}_c$. To simplify the following arguments, we will abuse notation and not distinguish between $H_0$ and $H_0$ restricted to a subspace of constant parity. At the end of this subsection, we will take the difference into account properly. 

Let us collect the Dirac fermion operators $c,c^\dagger$ into a $(2L+2)$-dimensional vector $\bc = (c_1, \dots, c_{L+1}, c_1^\dagger,\dots,c_{L+1}^\dagger)^t$. We express $H_0$ given by Eq.~\eqref{eq:H0_obc_c} as
\begin{equation}\label{eq:H0_obc_Mc}
    H_0 = \frac{1}{2} \bc^\dagger \begin{pmatrix}
        A &B \\ C &-A^t
    \end{pmatrix}
    \bc
    = \frac{1}{2} \bc^\dagger M_c \bc
\end{equation}
where the $(L+1)\times(L+1)$-matrices $A,B$ and $C$ are given by
\begin{align}
    A_{ij} &= -\delta_{i,j+1\mathrm{mod} (L+1)} \notag \\
    &+(1-\beta s)\delta_{i,L+1}\delta_{j,L} + (1-\alpha)\delta_{i,1}\delta_{j,L+1}, \label{eq:H0_obc_A}\\
    B_{ij} &= \alpha (\delta_{i,1}\delta_{j,L+1} - \delta_{i,L+1}\delta_{j,1}), \label{eq:H0_obc_B} \\
    C_{ij} &= \beta s(\delta_{i,L}\delta_{j,L+1} - \delta_{i,L+1}\delta_{j,L}), \label{eq:H0_obc_C}
\end{align}
and $\delta$ denotes the Kronecker-delta symbol. 

The matrix $A$ is, up to deformations in the $(1,L+1)$th and $(L+1,L)$th entries, a circulant matrix with only one non-zero off-diagonal. The matrices $B$ and $C$ only contain two non-zero entries. Thus, the solutions $\lambda$ and $u$ to the eigenvalue problem
\begin{equation}
    M_c u = \lambda u,
\end{equation}
are closely related to the eigen-decomposition of circulant matrices, which in turn are given by Fourier transforms. As shown in detail in Appendix~\ref{sec:app_Mc}, the eigenvalues $\lambda$ are solutions of
\begin{equation}\label{eq:H0_obc_roots}
    \lambda^{2L+2} = 4 (\alpha\beta)^2 (-1)^L,
\end{equation}
and are independent of the parity sector $s$. Since the polynomial in Eq.~\eqref{eq:H0_obc_roots} is of even degree, its roots appear in pairs of $\pm\lambda$.

The Hamiltonian $H_0$ in Eq.~\eqref{eq:H0_obc_c} is non-Hermitian, preventing the direct use of the (Hermitian) Bogoliubov-de-Gennes formalism for linking the eigenvalues of $M_c$ to the many-body spectrum of $H_0$. Hence, we will pursue an alternative method. We proceed as in \cite{Prosen_NewJournPhys_2008_third_quant} and express $c,c^\dagger$ in terms of Majorana fermions
\begin{equation}\label{eq:H0_obc_phi}
    \phi_{j,1} = \frac{1}{\sqrt{2}}(c_j+c_j^\dagger), \quad 
	\phi_{j,2} = \frac{1}{i\sqrt{2}}(c_j-c_j^\dagger).
\end{equation}
After collecting the Majorana fermions $\phi_{j,l}$ into a column vector $\bphi = (\phi_{1,1}, \phi_{1,2}, \dots, \phi_{L+1,1}, \phi_{L+1,2} )^t$, $H_0$ can be written as
\begin{align}
    H_0 = \frac{1}{2} \bphi^t M_\phi \bphi,
\end{align}
where the matrix $M_\phi$ is a complex and anti-symmetric $(2L+2)\times(2L+2)$-matrix. The transformation of Majorana fermions $\phi$ to Dirac fermions $c$ via Eq.~\eqref{eq:H0_obc_phi} is unitary, making $M_\phi$ and $M_c$ unitarily equivalent and hence sharing the same eigenvalues.

As $M_\phi$ is anti-symmetric, it can be factorized \cite{Prosen_NewJournPhys_2008_third_quant} as
\begin{equation}
    M_\phi = \frac{1}{2} V \Lambda J V^t
\end{equation}
where
\begin{equation}
    V^t V = J = \operatorname{Id}_{L+1} \otimes \begin{pmatrix}
		0 & 1 \\  1 & 0
	\end{pmatrix},
\end{equation}
$\operatorname{Id}_{L+1}$ denotes the $(L+1)\times(L+1)$ identity matrix and $\Lambda$ is a diagonal matrix containing the eigenvalues of $M_\phi$ ($M_c$). The anti-symmetry of $M_\phi$ implies that its eigenvalues come in pairs $\pm\lambda$, which is consistent with the solutions of Eq.~\eqref{eq:H0_obc_roots}. The diagonal of $\Lambda$ is ordered as $\lambda_1, - \lambda_1,\dots \lambda_{L+1}, -\lambda_{L+1}$. We fix the choice between $\lambda_j$ and $-\lambda_j$ by requiring $\Re \lambda_j \ge 0$.

Let us define another type of Dirac fermions $b, b'$ as 
\begin{equation}
    (b_1,b_1',\dots,b_{L+1},b_{L+1}')^t = \left( V^t \bphi\right).
\end{equation}
These fulfill the usual anti-commutation relations of Dirac fermions \cite{Prosen_NewJournPhys_2008_third_quant}, but $b'$ is in general not the Hermitian adjoint of $b$. Nevertheless, the Hamiltonian $H_0$ becomes diagonal in terms of $b,b'$,
\begin{equation}\label{eq:H0_obc_b}
    H_0 = \sum_{j=1}^{L+1} \lambda_j b_j'b_j - \frac{1}{2} \sum_{j=1}^{L+1} \lambda_j.
\end{equation}
The eigenstates of $H_0$ are given by creation operators $b_j'$ acting on the vacuum $\ket{0}_b$, which are $2^{L+1}$ in total. But not all eigenstates correspond to an eigenvalue of $H_0$ given by Eq.~\eqref{eq:H0_obc_c}. We have to take into account that the Dirac fermions $b,b'$ are only defined on fixed parity subspaces.

We numerically find that the parity operator $\mathcal{P}_b$ of the $b,b'$ fermions obeys
\begin{equation}
    \mathcal{P}_b = -s \mathcal{P}_c,
\end{equation}
where $P_c$ denotes the parity operator of the $c$ fermions. Recall that we let $s=(-1)^L \mathcal{P}_c$ at the beginning of this subsection. Thus, the admissible $b'$-fermion states must have $b$-parity $\mathcal{P}_b=-(-1)^L=(-1)^{L+1}$. Especially, the parity of the admissible $b$-states does not depend on $s$. Thus both parity sectors give rise to the same many-body spectrum of $H_0$ in Eq.~\eqref{eq:H0_obc_b}, as required.

In summary, the many-body spectrum of the non-interacting TASEP, subject to a global shift in the complex plane, is represented by the sums of the $L+1$ roots from Eq.~\eqref{eq:H0_obc_roots} with positive real parts. These are scaled roots of $\pm 1$ with magnitude proportional to $(\alpha\beta)^{1/(L+1)}$.  Depending on whether $L$ is odd or even, an even or odd number of summands, respectively, are included in the sums.

\subsection{Spectral boundary}\label{sec:H0_obc_boundary}

\begin{figure}
    \centering
    \includegraphics[width=1\linewidth]{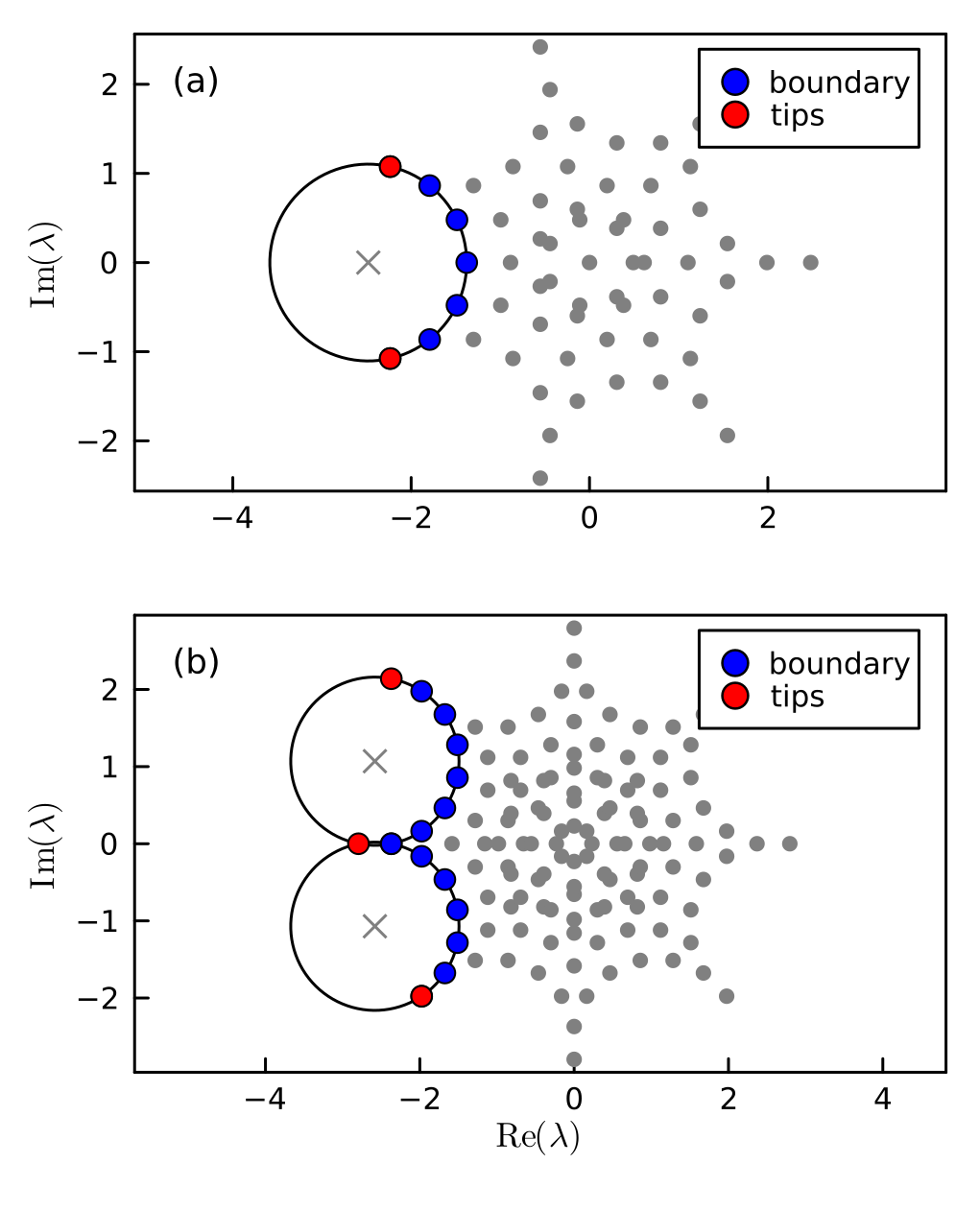}
    \caption{Many-body spectrum of the non-interacting TASEP with obc on \textbf{(a)} $L=6$ and \textbf{(b)} $L=7$ sites. Similar to pbc in Fig.~\ref{fig:H0_pbc}, all boundary eigenvalues lie on circles, with midpoints denoted by crosses. 
    %Highlighted boundary eigenvalues are given by Eq.~\eqref{eq:H0_obc_lambda_even} \textbf{(a)} and Eq.~\eqref{eq:H0_obc_lambda_odd} \textbf{(b)}
    }
    \label{fig:H0_obc}
\end{figure}

The emergence of the many-body spectrum of the non-interacting TASEP with obc follows a similar principle than for pbc discussed in Sec.~\ref{sec:H0_pbc}: the many-body spectrum consists of sums of (scaled) roots of $\pm1$. In the following, we describe how the spiky spectral boundary emerges for obc. Especially, we will demonstrate that, akin to the pbc case, the spectral boundary resides on $L+1$ circles, each with a radius of $(2\alpha\beta)^{1/(L+1)}$, and provide a comparable parametrization for this boundary.

In the following, we focus exclusively on the spectral boundary associated with the most negative real parts. This is illustrated in Fig.~\ref{fig:H0_obc}, parts (a) and (b), where the eigenvalues of the relevant sectors are marked with blue and red circles. The rotational symmetry of the spectrum means that the structure of the boundary is a repetitive pattern reflecting the shape of sectors with the smallest real parts. Hence, restriction to sectors with the most negative real part eigenvalues comes with no loss of generality.

Let us first consider even $L$. Recall that the many-body spectrum is given by sums of an odd number of positive real part roots of the polynomial in Eq.~\eqref{eq:H0_obc_roots}. Let us denote the $L+1$ roots with non-negative real part by $\lambda_1,\dots,\lambda_{L+1}$. Then the $L+1$ many-body eigenvalues with the smallest real parts are eigenvalues lying on the spectral boundary and given by
\begin{equation}\label{eq:H0_obc_lambda_even}
    \lambda_j - \frac{1}{2}\sum_{l=1}^{L+1} \lambda_l.
\end{equation}
If we label $\lambda_l$ by increasing angle with branch-cut on the negative imaginary axis then the tips of the spectrum are given by the indices $j=1$ and $j=L+1$.

In Fig.~\ref{fig:H0_obc}(a) we show the spectrum of the non-interacting TASEP with obc on $L=6$ sites. The spectrum shows $L+1=7$ spikes. The boundary and tips according to Eq.~\eqref{eq:H0_obc_lambda_even} are shown as blue and red markers, respectively. The markers lie on a circle with midpoint $-\frac{1}{2}\sum_{l=1}^{L+1} \lambda_l$ and radius $|\lambda_j| = (2\alpha\beta)^{1/(L+1)}$.

Let us now consider the slightly more complicated case of odd $L$. In Fig.~\ref{fig:H0_obc}(b) we show the many-body spectrum on $L=7$ sites. The tip of the spectral edge with the smallest real part is given by an `empty' sum of $\lambda_l$'s and thus is $-\frac{1}{2}\sum_{l=1}^{L+1} \lambda_l$. The boundary eigenvalues are given by the following (shifted) sum of two single-particle eigenvalues: 
\begin{equation}
    \lambda_j +\lambda_{1,L} -\frac{1}{2}\sum_{l=1}^{L+1} \lambda_l,
\end{equation}
where $2\le j \le L$ and $\lambda_1$ corresponds to the lower spectral boundary in Fig.~\ref{fig:H0_obc}(b) while $\lambda_L$ corresponds to the upper part. The midpoints of the circles are given by $\lambda_{1,L} -\frac{1}{2}\sum_{l=1}^{L+1} \lambda_l$ and the radius again by $|\lambda_j|= (2\alpha\beta)^{1/(L+1)}$.

Similar to the pbc case, we can establish a continuous boundary for the many-body spectrum, parametrized by Eq.~\eqref{eq:H0_boundary}. In the obc case the constants $\gamma_{1,2}$ are given by
\begin{align}
    \gamma_1 &= (2\alpha\beta)^{1/(L+1)} \frac{1}{2\sin(\pi/(2L+2))}\\
    \gamma_2 &= (2\alpha\beta)^{1/(L+1)},
\end{align}
while the piece-wise constant $f$ and the piece-wise identify function $g$ are given by
\begin{align}
    f(t) &= \frac{\pi}{L} \left( 2\left\lfloor \frac{Lt}{2\pi} \right\rfloor -1\right) \\
    g(t) &= \pi \frac{L+2}{2L+2} + \frac{L}{2L+2} (t \operatorname{mod}\ 2\pi).
\end{align}
The continuous boundary $z_B(t)$ with the above parameters is illustrated in Fig.~\ref{fig:spectrum_tasep}(d) as a red (gray in print) curve for $L=11$.

\subsection{Spikes in the large $L$ limit}\label{sec:H0_obc_spikes}
The parametrization of the spectral boundary for obc shows a clear link to the spectral boundary for pbc. Specifically, in the large $L$ limit with constant $\alpha, \beta$, the obc spectral boundary aligns with the pbc case at $\rho=1/2$. This relation is immediately evident for $\gamma_2$, $f$, and $g$. Further, a series expansion of $\gamma_1$ for large $L$ reveals that its leading term, $\gamma_1=L/\pi +O(1)$, is identical in both cases, with differences emerging only at $O(1)$.

Consequently, in the large $L$ limit, the spiky spectral boundary in the obc case remains pronounced. Rescaling the spectrum by $1/L$, the spectral density approaches filling the unit disk as $L\to\infty$. For finite $L$, the tips are spaced by $O(1/L)$, and the maximum deviation of the boundary from the unit circle is also $O(1/L)$.

%%%%%%%%%%%%%%%%%%%%%%%%%%%%%%%%%%%%%%%%%%%%%%%%%%%%%%%%%%%%%%%%%%%%%%%%%%%%%%%%%%%%%
%%%%%%%%%%%%%%%%%%%%%%%%%%%%%%%%%%%%%%%%%%%%%%%%%%%%%%%%%%%%%%%%%%%%%%%%%%%%%%%%%%%%%
%%%%%%%%%%%%%%%%%%%%%%%%%%%%%%%%%%%%%%%%%%%%%%%%%%%%%%%%%%%%%%%%%%%%%%%%%%%%%%%%%%%%%
\section{Pbc TASEP by Bethe ansatz}\label{sec:bethe}

In Sec.~\ref{sec:H0_pbc}, we showed that in the non-interacting TASEP ($U=0$) with pbc, the spiky boundary of the many-body spectrum emerges essentially as sums of evenly spaced single-body eigenvalues $\lambda_1,\dots,\lambda_L$. This section expands that concept to interaction strengths $0<U$. Employing the coordinate Bethe ansatz, we generalize the single-body framework to Bethe roots, which tend to cluster close to $\lambda_1,\dots,\lambda_L$. This clustering, combined with TASEP many-body eigenvalues being sums of Bethe roots, results in a spiky spectrum boundary for any interaction strength $0\le U\le1$.

This section focuses on $\rho\approx 1/2$, where the most prominent spectral boundary spikes in the non-interacting ASEP were observed. In the low-density limit ($\rho$ approaching zero), we anticipate a spectral boundary for the usual ASEP similar to the non-interacting case, characterized by a smooth, circular boundary without spikes. Fig.~\ref{fig:spectrum_tasep_small_N} partly supports this, showing similar many-body spectra for TASEP with $U=1$ (a) and $U=0$ (b), both featuring smooth, non-spiky spectral boundaries.

Sec.~\ref{sec:bethe_ansatz} generalizes the coordinate Bethe ansatz to arbitrary $U$, with derivation details and numerical solution methods detailed in Appendices~\ref{sec:app_bethe} and \ref{sec:app_bethe_numerical}. In Sec.~\ref{sec:bethe_roots}, we demonstrate the clustering of solutions to the Bethe equations and in Sec.~\ref{sec:bethe_many_body}, we establish how this clustering results in a spiky spectral boundary.

\subsection{Coordinate Bethe ansatz}\label{sec:bethe_ansatz}
We start by determining the many-body spectrum of $H$, as described in Eqs.~\eqref{eq:H0_pbc_def} and \eqref{eq:I_pbc_def}, for arbitrary $U$. We closely follow the application of the coordinate Bethe ansatz to the ASEP in \cite{Gwa_Spohn_PRA1992_Bethe_ansatz}, which dealt with $U=1$.   The coordinate Bethe ansatz has since been used extensively for ASEP 
\cite{Dhar_PhaseTrans1987_asep, Gwa_Spohn_PRA1992_Bethe_ansatz, Kim_PRE1995_xxz_chain_bethe_ansatz, Derrida_PhysRep1998ASEP, Derrida_Lebowitz_PRL1998_dev_functions_asep, Golinelli_Mallick_JourPhysA2004_asep_spectral_gap,
Golinelli_Mallick_JourStatMech2004_hidden_symmetry,
Golinelli_Mallick_JourPhysA2005_gap_tasep,
Golinelli_Mallick_JourStatMech2005_tasep_degeneracies,
Golinelli_Mallick_JourPhysA2006ASEP, 
Appert-rolland_Derrida_etal_PRE2008_cumulants, Prolhac_Mallick_JPhysA2008_CurrentFluc,
Prolhac_JourPhysA2010_tree_structure,
Mallick_JSTAT2011_SomeExactResults,
Chou_Mallick_Zia_RepProgPhys2011,
Simon_JourStatPhys2011_weakly_asep,
Motegi_etal_PRE_2012_bethe_dynamics,  Prolhac_JourPhysA2013_bulk_evs,
Corwin_2014_coord_bethe,
Prolhac_JourPhysA2014_first_excited,
Prolhac_JournStatMech2015_fluct_dev_TASEP, 
Mallick_PhysicaA2015_ASEP,
Prolhac_PRL2016, Prolhac_JPhysA2016_ExtrapolationMethods, Prolhac_JPhysA2017_WeaklyAsymmetric, Prolhac_JPhysA2020_RiemannSurfacePBC, LiuSaenzWang_CommMathPhys2020_IntegralFormulas, Prolhac_JPhysA2021_RiemannSurface, Baik_Liu_ProbTheo2021_tasep}
and for various variants/extensions of ASEP 
\cite{Noh_Kim_PRE1994_five_vertex, Schutz_JourStatPhys1997_infinite_asep, Alimohammadi_PRE1998_gen_asep, Sasamoto_Wadati_PRE1998_no_exclusion, Derrida_Evans_JourPhysA_1999_defect, Alcaraz_Bariev_PRE1999_any_size_particles, Karimipour_EPL1999_bethe_ansatz_mps, Roshani_Khorrami_PRE1999_cont_asep, Roshani_Khorrami_PRE2001_multi_species,Alcaraz_Bariev_PhysicaA2002_higher_spin, Ferreira_Alcaraz_PRE2002_any_size_particles, Roshani_Khorrami_JourMathPhys2002_gen_asep,
Priezzhev_PRL2003_asep, Povolotsky_etal_PhysicaA2003_avalanche_process, Priezzhev_PRL2003_tasep_prob, Povolotsky_PRE2004_zerorange, Roshani_Khorrami_EPJB2005_discrete_time, Golinelli_Mallick_JourPhysA2006ASEP, Povolotsky_Mendes_JourStatPhys2006_parallel_update, Povolotsky_Priezzhev_JourStatMech2006_parallel_update, Povolotsky_Priezzhev_JourStatMech2007_parallel_update_2, Majumdar_etal_PRE2008_5vertex, 
Simon_JourStatMech_2009_obc_coord_bethe, Lazo_Ferreira_PRE2010_impurity,Lazo_Ferreira_JSTAT2012_more_impurities, Lee_JourStatPhys2012_mulitparticle, Borodin_etal_CommMathPhys2015_coord_bethe, Derbyshev_etal_PRE2015_gen_asep, LiuSaenzWang_CommMathPhys2020_IntegralFormulas, Chen_deGier_etal_CommMathPhys2022_two_species, Saenz_Tracy_Widom_book_2022, deG_etal_JourPhysA2023_multispecies, Ishiguro_etal_PRR2023_nonHermtian_spin_chain, Lobaskin_Evans_Mallick_JourPhysA2023_two_species}.  
However, to the best of our knowledge an extension to general $U$ has not been presented before; we present this extension in this work.  The details of the  derivation of the Bethe equations are provided in Appendix~\ref{sec:app_bethe}; in this section we present the results and their application to understanding the spectral boundary.

The Bethe ansatz eigenvalues $E$ for arbitrary $U$ are given by
\begin{equation} \label{eq:bethe_energy}
E = \sum_{j=1}^N \big( pz_j + qz_j^{-1} - U \big),
\end{equation}
where $z_j$ are complex numbers, the so-called Bethe roots, which in turn are solutions of the following recurrent relations
\begin{equation}\label{eq:bethe_recurrent}
	z_j^L = \prod_{k=1;k\neq j}^N \left( - \frac{q+pz_jz_k-U z_j}{q+pz_jz_k-U z_k} \right).
\end{equation}
The solutions of Eq.~\eqref{eq:bethe_recurrent} are $N$-tuples $(z_1,\dots,z_N)$ and each $N$-tuple gives rise to an eigenvalue $E$ of the TASEP $H$ via Eq.~\eqref{eq:bethe_energy}. 

Numerical data indicates that in small systems, each eigenvalue is a sum of Bethe roots, although a formal proof of the completeness of the Bethe ansatz is lacking  \cite{Dorlas_CommMathPhys1993_Bethe_completeness, Langlands_Saint-Aubin_1995_algebro-geometric, Brattain_Do_Saenz_arxiv2017_bethe_ansatz_complete}.  In our finite ASEP system investigations, all eigenvalues conformed to the Bethe ansatz.

For $p=1$ and $q=0$, the eigenvalue equation simplifies to
\begin{equation}\label{eq:bethe_tasep_energy}
E = \frac{1}{2}\sum_{j=1}^N \big(Z_j - U \big),
\end{equation}
and the Bethe equations transform into
\begin{equation}\label{eq:bethe_tasep_recurrent}
\left(U+Z_j\right)^{L-N}\left(U-Z_j\right)^N = - 2^L \prod_{k=1}^N \frac{Z_k-U}{Z_k+U},
\end{equation}
with 
\begin{equation}
Z_k=2z_k-U
\end{equation}
representing scaled, shifted Bethe roots.  We refer to either the $z_k$ or the $Z_k$ as the Bethe roots, depending on the context.  
In Eqs.~\eqref{eq:bethe_tasep_recurrent}, the main simplification from the general $p,q$ case is the independence of the right-hand side from $j$, which makes the solutions $Z_j$ roots of the polynomial $P(Z) = (U+Z)^{L-N}(U-Z)^N - Y$, with $Y$ given by the right-hand side of Eq.~\eqref{eq:bethe_tasep_recurrent}. This not only simplifies numerical computation of the Bethe roots $Z_j$, but also ensures their continuity in $U$ \cite{Brillinger_MathMag_1966_pol_analytic}. Consequently, we will focus on the specific case of $p=1$ and $q=0$ for the rest of this section.

Appendix~\ref{sec:app_bethe_numerical} details the numerical solution process for the Bethe Eqs.~\eqref{eq:bethe_tasep_recurrent} and the systematic retrieval of all Bethe roots.

Eq.~\eqref{eq:bethe_energy} and \eqref{eq:bethe_tasep_energy} establish that many-body eigenvalues are sums of Bethe roots, up to a global shift. To demonstrate a spiky spectral boundary, we will show numerically a sufficient clustering of Bethe roots, which is the focus of the rest of this section.

\subsection{Clustering of the Bethe roots}\label{sec:bethe_roots}

\begin{figure}
	\begin{center}
		\includegraphics[width=\columnwidth]{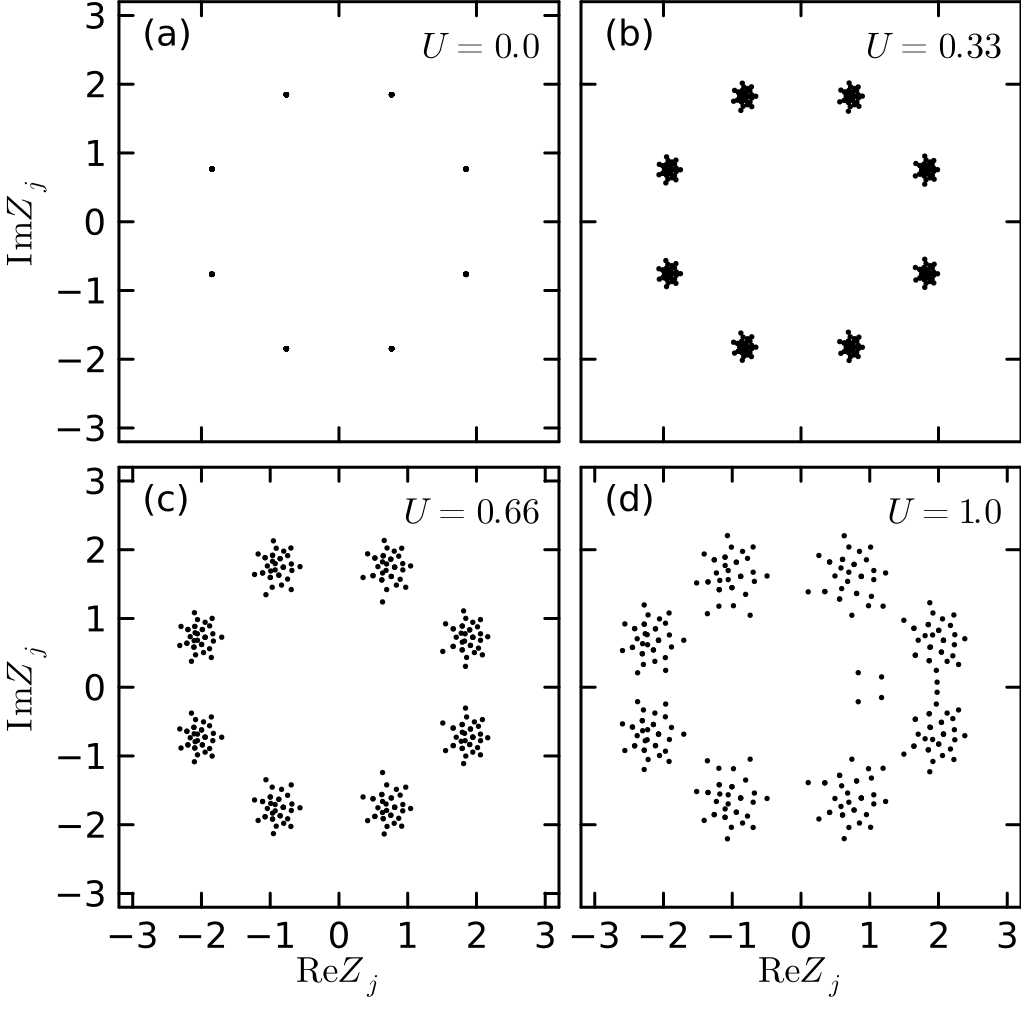}
        \includegraphics[width=\columnwidth]{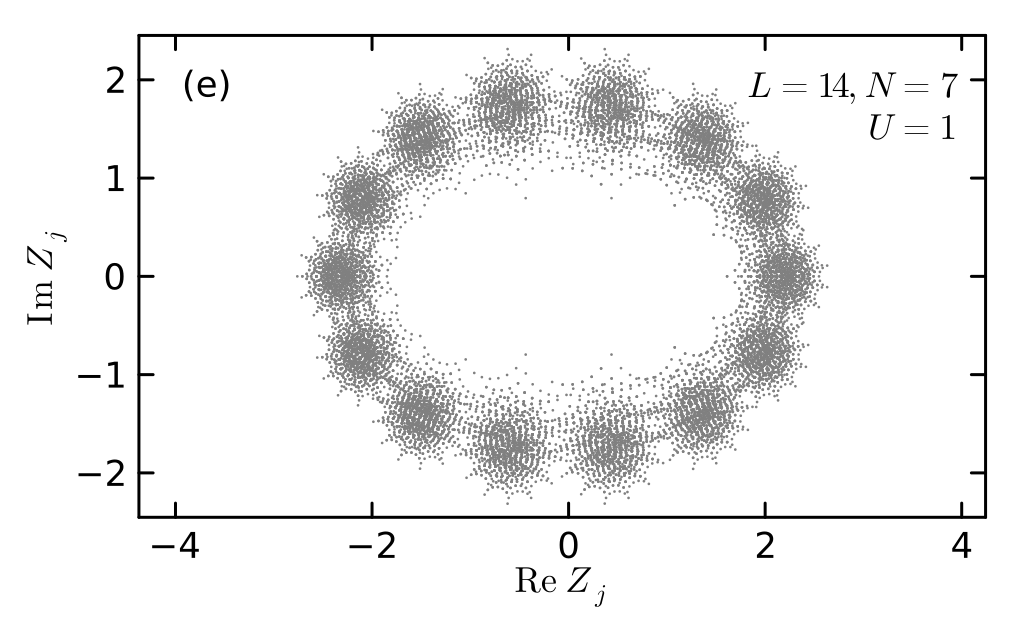}
		\caption{All $N\cdot\binom{L}{N}$ Bethe roots $Z_j$ of the TASEP.  \textbf{(a)-(d)} $L=8$, $N=4$, for different values of $U$. \textbf{(e)} $L=14$, $N=7$,  $U=1$. \label{fig:bethe_roots}}
	\end{center}
\end{figure}

To examine the spectral boundary in terms of the Bethe roots, we will consider in the complex plane the Bethe roots ($z_j$ or $Z_j$) corresponding to each of the $\binom{L}{N}$ eigenstates.  There are thus $N\times\binom{L}{N}$ Bethe roots in total, for any value of $U$.  Such plots are shown in Figure~\ref{fig:bethe_roots}.  

For $U=0$, the Bethe roots $z_j$ satisfy the equation $z_j^L = (Z_j/2)^L = (-1)^{N+1}$, and agree with the single-body eigenvalues of $H_0$ as stated in Eq.~\eqref{eq:H0_pbc_sb}. Therefore, the many-body spectrum derived via the Bethe ansatz for $U=0$ aligns with that of the non-interacting ASEP model discussed in Section~\ref{sec:H0_pbc}, as expected. An illustrative example of the Bethe roots $Z_j=2z_j$ for $U=0$ is provided in Figure~\ref{fig:bethe_roots} (a) for $L=8$ and $N=4$.  Here, each solution of the Bethe equations contributes $N=4$ roots, which together describe one of the $\binom{8}{4}$ eigenstates.  We plot all the $4\times\binom{8}{4}$ roots together in a single plot.  Since for $U=0$ every solution to the Bethe equations is a subset of the 8 single-body eigenvalues of $H_0$, the union of all solutions is highly degenerate and only 8 unique markers show up in Fig.~\ref{fig:bethe_roots}(a).

For $U>0$ the degeneracy of the $U=0$ case is lifted and the $4\times\binom{8}{4}$ Bethe roots $Z_j$ become distinct, as observed in Fig.~\ref{fig:bethe_roots}(b-d) for $U=0.33$, $0.66$, and $U=1$, respectively. The continuity of Bethe roots $z_j$ in $U$ suggests that for small $U$, these roots should be proximate to the $L$th roots of $(-1)^{N+1}$. Numerically, this is confirmed as the Bethe roots $z_j$ tend to cluster around the $L$th roots of $(-1)^{N+1}$ for small $U$. As depicted in Figure~\ref{fig:bethe_roots} (b) and (c) for $U=0.33$ and $U=0.66$ respectively, the $Z_j$'s distinctly form $L=8$ clusters around the Bethe roots for $U=0$. This clustering is even discernible for $U=1$, as shown in Figure~\ref{fig:bethe_roots} (d), where the $L=8$ clusters remain identifiable.

For larger $L$, the Bethe root clusters overlap at $U=1$, evident from Fig.~\ref{fig:bethe_roots}(e) for $L=14$ and $N=7$. However, the statistical width of these clusters diminishes with larger $L$. This is demonstrated in Fig.~\ref{fig:stds_vs_L}, where the average cluster width decreases as $L^{-1/2}$ in the thermodynamic limit with $\rho=N/L=1/2$ and $N,L\to\infty$.

We define the locations and widths of these clusters by fitting a Gaussian mixture model of $L$ independent Gaussians $\mathcal{N}$ with complex means to the Bethe roots. The Bethe roots distribution is approximated as $\frac{1}{L} \sum_{j=1}^L f_j$, with $f_j$ representing Gaussian densities.  We label the Gaussians of the optimal fit as $\mathcal{N}_j$, each characterized by its mean $\mu_j$ and standard deviation $\sigma_j$.

\subsection{Structure of the many-body spectrum}\label{sec:bethe_many_body}

\begin{figure}
	\begin{center}
		\includegraphics[width=\columnwidth]{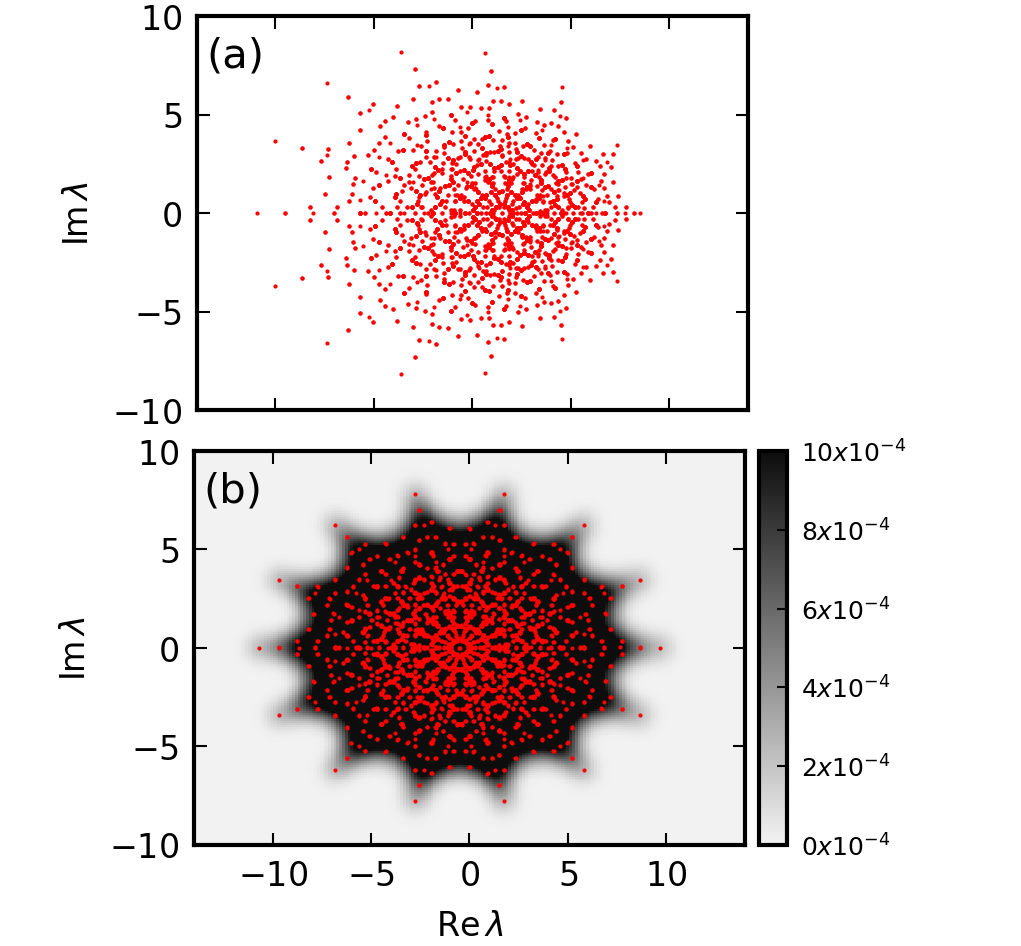}
		\caption{\textbf{(a)} The many-body spectrum of TASEP with $L=14$ and $N=7$ (multiplied by 2 and shifted by $N$). \textbf{(b)} Probability density function of the many-body spectrum of the random Bethe roots $Z$ for $L=14$ and $N=7$ capped at $10^{-3}$. Red (gray in print) dots are the means of the complex Gaussians. \label{fig:bethe_random_mb}}
	\end{center}
\end{figure}

In the following, we will show that by considering only the centers and widths of Bethe root clusters, and not their specific structure, we can approximate a many-body spectrum that mirrors key characteristics of the TASEP many-body spectrum, particularly its spiky boundary.

Recall that for $U=0$ each many-body eigenvalue $E$ is a sum of $N$ out of $L$ single-body eigenvalues. Specifically, $E$ is given by
\begin{equation}
    E = \sum_{j=1}^L s_j \lambda_j = \sum_{s_j\neq 0} \lambda_j,
\end{equation}
where $s\in \{0,1\}^L$ is a configuration with $\sum_j s_j = N$ and $\lambda_j$ are the single-particle eigenvalues determined in Sec.~\ref{sec:H0_pbc_boundary}. By Eq.~\eqref{eq:bethe_tasep_energy} every many-body eigenvalue of the TASEP ($U=1$) corresponds to a sum of $N$ Bethe roots $(Z_1,\dots,Z_N)$ and by the continuation from $U=1$ to $U=0$ each Bethe root $Z_j$ belongs to one of the $L$ clusters. Instead of summing solutions of the Bethe Eqs.~\eqref{eq:bethe_tasep_recurrent} we employ a statistical ansatz and consider random many-body eigenvalues of the form
\begin{equation}
    E_{rand} = \frac{1}{L}\sum_{j=1}^L s_j \mathcal{N}_j = \mathcal{N}_s,
\end{equation}
where $\mathcal{N}_s$ denotes a Gaussian with mean $\sum_{j=1}^L s_j\mu_j$ and variance $\sum_{j=1}^L s_j\sigma_j^2$. We refer to $\mathcal{N}_s$ as many-body Gaussians. The full random many-body spectrum is then given by
\begin{equation}\label{eq:bethe_tasep_random_mb}
    \frac{1}{\mathcal{Z}} \sum_{\substack{s\in\{0,1\}^L \\ s_1+\dots+s_L = N}} \mathcal{N}_{s}
\end{equation}
where $\mathcal{Z}=L\binom{L}{N}$ is a normalization constant. Keep in mind that the Gaussians $\mathcal{N}_{s}$ for different configurations $s$ are independent. The many-body spectrum of the TASEP is a specific sample of the distribution in Eq.~\eqref{eq:bethe_tasep_random_mb}. For $U=0$ the random spectrum becomes deterministic and agrees with the non-interacting many-body spectrum presented in Sec.~\ref{sec:H0_pbc}.

In Fig.~\ref{fig:bethe_random_mb}(b), we present the probability density from Eq.\eqref{eq:bethe_tasep_random_mb} for $L=14$, $N=7$, and $U=1$, with the density capped at $10^{-3}$ for clarity. The red markers indicate the means $\sum_{j=1}^L s_j\mu_j$ of the many-body Gaussians $\mathcal{N}_s$. Both the discrete means and the continuous density exhibit pronounced spikes at the boundary. When these means are compared to the TASEP many-body spectrum shown in Fig.~\ref{fig:bethe_random_mb}(a), even finer details of the spectrum are discernible in the structure of the means.

The boundary of the random many-body spectrum is mainly determined by Gaussians $\mathcal{N}_s$, associated with domain wall configurations of one or two domain walls, separated by at most one empty site, due to the exponential decay of the Gaussian probability density function. These configurations are identical to those defining the spectral boundary in the non-interacting case.

The random Bethe spectrum and the TASEP spectrum share a remarkably similar overall shape. However, differences do exist, e.g., the boundary of the random Bethe spectrum is not  skewed leftwards in the complex plane. This is attributed to the additional structure in the Bethe root clusters seen in Fig.~\ref{fig:bethe_roots}, not represented by rotationally invariant Gaussians.

\subsection{``Thermodynamic limit''}

\begin{figure}
	\begin{center}
		\includegraphics[width=\columnwidth]{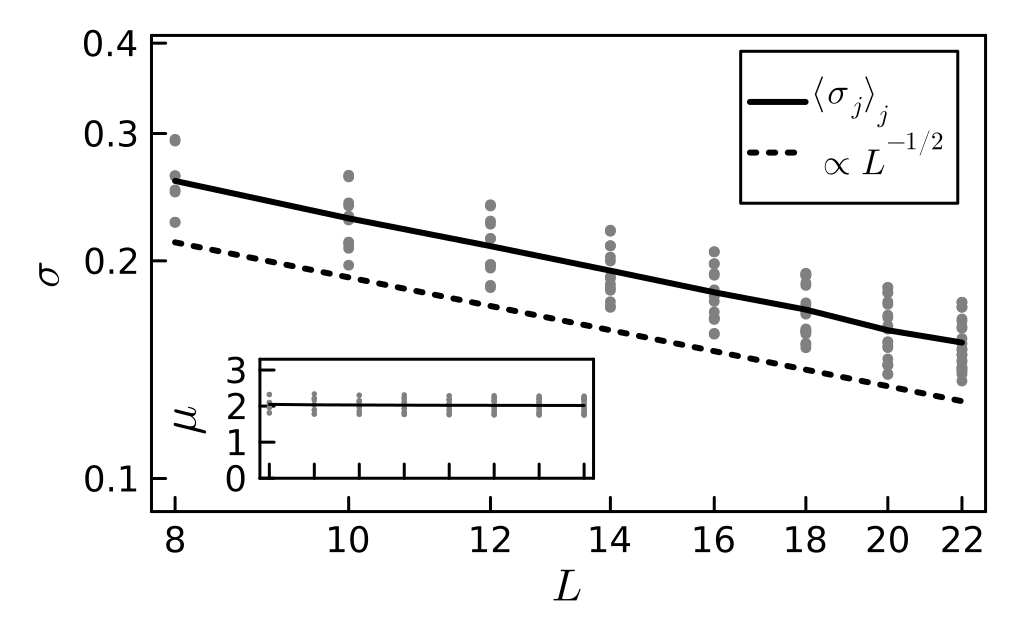}
		\caption{The width $\sigma$ of the complex Gaussians fitted to the clusters of the Bethe roots for $U=1$ at half-filling $N=L/2$. The solid line denotes the average $\langle \sigma_j\rangle_j = \frac{1}{L}\sum_j \sigma_j$ of the cluster widths and the dotted line guides the eye to $L^{-1/2}$. The \textbf{inset} shows the absolute value of the centers of the complex Gaussians $|\mu|$. Black solid line indicates the average. \label{fig:stds_vs_L}}
	\end{center}
\end{figure}

Similar to the non-interacting case with $U=0$, we demonstrate that the spiky boundary persists in the thermodynamic limit as $L$ and $N$ increase while maintaining a fixed density $\rho=N/L$.

Let us first focus on the centers $\sum_{j=1}^L s_j \mu_j$ of the many-body Gaussians $\mathcal{N_s}$, depicted as red dots in Fig.~\ref{fig:bethe_random_mb}. According to the inset of Fig.~\ref{fig:stds_vs_L}, the absolute values of $|\mu_j|$ appear to be independent of $L$. This independence suggests that the non-interacting case scenario also applies to the many-body Gaussian centers. For boundary configurations $s$, these centers, being sums of $N=\rho L$ nearby $\mu_j$, scale with $L$. Given that both the tip distance ($d_t$ from Sec.\ref{sec:H0_pbc_spikes}) and boundary depth ($d_b$ from Sec.\ref{sec:H0_pbc_spikes}) are proportional to 1, the spiky structure of the boundary Gaussian centers is maintained in the thermodynamic limit.

However, this does not automatically mean that the spiky spectral boundary of the random spectrum, as defined in Eq.~\eqref{eq:bethe_tasep_random_mb}, persists in the thermodynamic limit. For this to hold true, the widths of the Gaussians $\mathcal{N}_j$ in the mixture model must decrease sufficiently fast.

Fig.~\ref{fig:stds_vs_L} displays the widths $\sigma_j$ of $\mathcal{N}_j$ for the TASEP case ($U=1$) at half-filling ($N=L/2$), with $L$ ranging from 8 to 22. The cluster widths $\sigma_j$ vary, being larger for clusters with smaller $|\Re Z|$ and smaller for those with larger $|\Re Z|$, as also observed in Fig.~\ref{fig:bethe_roots}(e). Despite this variation, the widths $\sigma_j$ are centered around their average $\langle \sigma_j\rangle_j = \frac{1}{L}\sum_{j=1}^L \sigma_j$, which decreases approximately as $\propto L^{-1/2}$, as shown by the dashed line in Fig.~\ref{fig:stds_vs_L}. Consequently, the variance $\sigma_s= \sum_{j=1}^L s_j\sigma_j^2$ of the Gaussians $\mathcal{N}_s$ scales as $\propto 1$. This indicates that the standard deviation of the boundary Gaussians $\mathcal{N}_s$ remains on the order of $\propto 1$ even as $L$ increases, aligning with the scale of both the tip distance and spike depths. Therefore, the spiky structure of the statistical many-body spectrum for $U=1$ is preserved in the thermodynamic limit, as in the $U=0$ case presented in Sec.~\ref{sec:H0_pbc}.

\section{The random matrix picture}\label{sec:random_graph}

\begin{figure}
    \centering
    \includegraphics{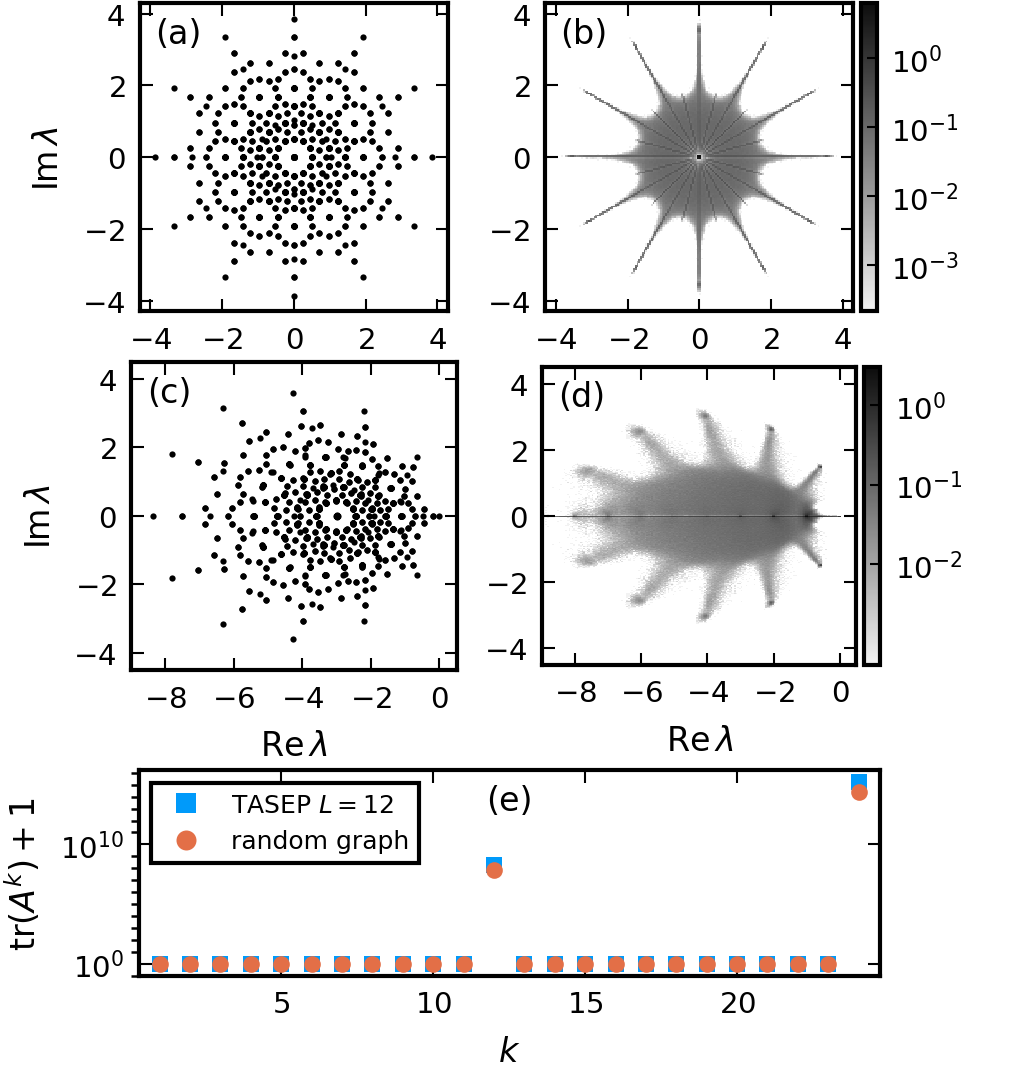}
    \caption{TASEP spectrum (pbc) with $L=12$ and $N=6$ for (a) $U=0$ and (c) $U=1$. In (b,d) spectral density of random graphs with cycle length divisible by $L$; in (b) of the adjacency matrix and in (d) of the negative (combinatorial) Laplacian. In (e) traces of powers of the non-interacting TASEP generator $H_0$ (squares) and random graph adjacency matrix (circles).}
    \label{fig:random_graphs}
\end{figure}

In the previous sections, we showed that the spikes of the spectral boundary of the TASEP are a consequence of the many-body spectrum being generated by summing single-particle-like clusters.

This section demonstrates that the spiky spectral boundary is a prevalent characteristic in a broad range of systems, extending beyond free fermions or those solvable by the Bethe ansatz. Specifically, this feature is typical in systems where the many-body graph exhibits a particular cycle structure, with cycle lengths being integer multiples of the spike count.

\subsection{From TASEP to graphs}

The matrix elements of the generator of the non-interacting TASEP $H_0$ are either zero or one. Thus the generator matrix is naturally interpreted as the adjacency matrix of a directed graph. This graph, which we will call the many-body graph of TASEP, has vertices representing particle configurations in the chain and edges indicating permissible transitions. For TASEP with $U=1$, its generator matrix $H$ is the negative combinatorial Laplacian of this graph.

\subsection{Cycles of TASEP}

The permissible transitions between particle configurations impose constraints on the structure of the many-body graph. Our focus is on the nature of cycles in the many-body graph, which are closed walks with only the start and end vertices being the same.

The cycle lengths in the TASEP many-body graph are divisible by $L$ for pbc and by $L+1$ for obc \cite{Prolhac_JourPhysA2013_bulk_evs}. This is evident in cycles among configurations, which only contain a single particle. These cycles consist of $L$ particle movements ($L+1$ for obc) such that the particle arrives at its original position.

The number of closed walks with length $k$ is related to entries of the $k$th power of the adjacency matrix $A$ ($A=H_0$ in the case of ASEP). The element $(A^k)_{ij}$ denotes the number of distinct walks of length $k$ from vertex $i$ to $j$. Thus $(A^k)_{ii}$ counts the number of distinct closed walks with length $k$ starting and ending at vertex $i$ and $\tr(A^k)$ aggregates the total number of closed walks with length $k$. Especially, if $\tr(A^k)=0$ then the graph does not contain any closed walks, thus any cycle, of length $k$.

In Fig.~\ref{fig:random_graphs}(e), we depict $\tr(A^k)+1$ as blue squares, where $A=H_0$, plotted against $k=1,\dots,2L$ for a system of $L=12$ sites and pbc with $N=6$ particles. The addition of $+1$ facilitates a logarithmic scale on the y-axis. Here, $\tr(A^k)$ equals zero for all values of $k$ not divisible by $L$, indicating the absence of cycles in the graph with length $k\ \mathrm{mod}\ L \neq 0$. Similarly, for obc, $\tr(A)^k=0$ if and only if $k\ \operatorname{mod}\ L+1 =0$ (not shown).

\subsection{Random graph model}
To demonstrate the robustness of the spiky spectral boundary, we compare the TASEP spectrum with the spectral density of a random graph ensemble characterized only by cycles whose lengths are divisible by $L$. This comparison is focused on the TASEP with pbc, noting that the obc scenario can be similarly analyzed by simply adjusting $L$ to $L+1$.

We sample the random graph by initially forming a directed cycle with $D$ vertices. Next, we randomly choose a vertex and traverse the graph randomly for $L-1$ steps. The vertex reached after $L-1$ steps is connected back to the starting vertex, creating a cycle of length $L$. This process is repeated until the graph contains a predetermined total number $n$ of edges.

Typically, the longest closed walk in the graph is the initial directed cycle linking all $D$ vertices. When the number of vertices $D$ is divisible by $L$, the construction of the graph ensures that all cycle lengths in the random graph are also divisible by $L$.

Fig.~\ref{fig:random_graphs} contrasts the random graph ensemble to the TASEP with $L=12$ sites and pbc with $N=6$ particles. Quantities of the random graph ensemble are averaged over $2,000$ samples, with the cycle length set to $L$ and the number of vertices $D=924$, matching the Hilbert space dimension of the TASEP.

In Fig.~\ref{fig:random_graphs}(e) we present $\tr(A^k)+1$ for the random graph ensemble, shown as red circles. In this ensemble, $\tr(A^k)$ is zero for all $k$ that are not integer multiples of $L$. Whenever $k$ is an integer multiple of $L$, $\tr(A^k)$ for the adjacency matrix $A$ of the random graph ensemble is comparable in magnitude to $\tr(A^k)$ for $A=H_0$, the generator matrix of TASEP. This similarity suggests that the number of closed walks in the random graph ensemble is on par with that in the TASEP many-body graph.

Fig.~\ref{fig:random_graphs}(a-d) displays a comparison between the random graph ensemble and TASEP, matching the parameters used in (e). In (a) and (b), we show the non-interacting TASEP spectrum alongside the estimated spectral density of the graph ensemble - both featuring $L$ distinct spikes.

In Fig.~\ref{fig:random_graphs}(c) and~(d), the focus is on the spectrum of TASEP ($U=1$) and the spectral density of the negative graph Laplacian for the random graph ensemble. Notably, the random graph Laplacian also presents $L$ pronounced spikes. The spike patterns, particularly their "bending" towards the left, show a resemblance to the TASEP spikes. The overall shape of the spectral density (ignoring the spikes) takes on a spindle-like form, characteristic of (sparse) random Markov matrices \cite{Timm_PRE_2009, Denisov_et_al_PRE_2019, Tarnowski_Denisov_et_al_PRE_2021, Nakerst_Denisov_Haque_PRE_2023}.

\section{Conclusion and Discussion}\label{sec:discussion}

In this work, we explored the connections among the spectral problems for ASEP, free fermion models, and random matrix theory, focusing particularly on the distinctive spiky shape of the ASEP spectral boundary.  We reformulated the ASEP generator matrices as non-Hermitian fermionic models with a variable interaction parameter $U$, where $U=1$ corresponds to the standard ASEP. We analytically demonstrated that in the non-interacting ASEP ($U=0$), this spiky spectral boundary arises from aggregating single-particle eigenvalues positioned on ellipses (circles for TASEP). For pbc, we extended this concept to interacting TASEP, showing that the spiky boundary remains and originates from the summation of clustered Bethe roots. Lastly, we confirmed the robustness of this spiky boundary by considering only the cycle structure in the many-body graph, revealing that corresponding random graphs exhibit a similar spiky spectral boundary.

This research opens up several questions for further exploration. We demonstrated the spiky spectral boundary in TASEP, largely attributed to Bethe roots clustering. It is intriguing to consider whether such clustering also occurs in ASEP. The straightforward connection between TASEP and ASEP in their non-interacting forms suggests that the spiky spectral boundary might extend to standard ASEP (with $U=1$) as well. However, it remains to be seen how introducing interactions influences Bethe roots clustering and the potential emergence of a spiky spectral boundary.

In this study, we concentrated on the Bethe ansatz for pbc. The ASEP with obc is also solvable via the Bethe ansatz, though the equations are more complex, as detailed in various studies \cite{DeGier_Essler_PRL_2005, DeGier_Essler_JSTAT2006, Simon_JourStatMech_2009_obc_coord_bethe, deGier_Finn_Sorrell_JPhysA2011,  Crampe_Ragoucy_Simon_JourStatMech_2011_obc_coord_bethe}.  One might ask whether the spiky spectral boundary in the obc case is also associated with a clustering of Bethe roots similar to the pbc case.

The spectral boundary of random graphs with dominant cycle lengths typically follows a hypotrochoidic curve, as noted in \cite{Aceituno_Rogers_Schomerus_PRE2019}. These graphs usually lack cycles shorter than $L$ but can have cycles longer than $L$. However, the random graph ensemble we introduced deviates from this standard hypotrochoidic pattern, likely due to its more restricted cycle structure, where all cycles are of lengths divisible by $L$. Extending the hypotrochoidic law to encompass this specific graph ensemble would be a valuable advancement.

This study concentrated on the spiky spectral boundary of the ASEP. Formation of spikes has as well been observed in the off-diagonals of reduced density matrices in the symmetric simple exclusion process (XXX model) \cite{Alba_PRB_2015} and the observable representation of Ising chain Glauber dynamics \cite{Gaveau_Schulman_JournPhysA_2006}. These observations together with the robustness of the spiky spectral boundary to perturbations make the the investigation of other models, both classical and quantum, that possess a similar cycle structure in their many-body graphs or comparable trace correlations in their generator matrices, an intriguing direction for future research.

\acknowledgements
GN and MH thank R. Taggart and P.C. Burke and GN thanks T. Giamarchi and I. Lobaskin for helpful discussions. This research is supported by the Deutsche Forschungsgemeinschaft through SFB No. 1143 (Project ID No. 247310070) (GN and MH) and the Irish Research Council Government of Ireland Postgraduate Scholarship Scheme (GOIPG/2019/58) (GN). TP is supported by the Grants N1-0219, N1-0334 as well as Program P1-0402 of Slovenian Research and Innovation Agency (ARIS).

\appendix

\section{Quadratic Fermion Model for obc}\label{sec:app_H0_obc}
In this section, we will show that the non-interacting ASEP $H_0$ with obc is a quadratic fermion model. Especially, we will prove Eq.~\eqref{eq:H0_obc_c}.

Recall
\begin{equation}
	H_0 = \sum_{j=1}^{L-1} \left(p\sigma_{j+1}^+ \sigma_j^- + q\sigma_j^+ \sigma_{j+1}^- \right) + \alpha \sigma_1^+ + \gamma \sigma_1^- + \beta \sigma_L^- + \delta \sigma_L^+,
\end{equation}

For that, we first apply, as mentioned in the main text, the Kramers-Wannier duality transformation \cite{Kogut_RevModPhys_1979}
\begin{equation}
    \sigma_j^x \to \prod_{l=1}^j \sigma_l^z,\quad
	\sigma_j^z \to \sigma_j^x \sigma_{j+1}^x,
\end{equation}
where we implicitly have enlarged the chain of length $L$ by one additional site to a chain of length $L+1$. Thus the multiplicity of every eigenvalue of the so-transformed $H_0$ is doubled. Applying a Jordan-Wigner transformation
\begin{equation}
	w_j = \left( \prod_{l=1}^{j-1} \sigma_l^z \right) \sigma_j^-, \quad
	w_j^\dagger = \left( \prod_{l=1}^{j-1} \sigma_l^z \right) \sigma_j^+,
\end{equation}
and rewriting in terms of Majorana ``real'' and ``imaginary'' parts of the Dirac fermions $w, w^\dagger$,
\begin{equation}
    \gamma_{j,1} = w_j^\dagger + w_j,\quad
	\gamma_{j,2} = i(w_j^\dagger - w_j),
\end{equation}
the Hamiltonian $H_0$ is given by
\begin{widetext}
\begin{align}
	H_0 =  &\sum_{j=1}^{L-1} \left[ \frac{p+q}{4}\left(i\gamma_{j+1,1}\gamma_{j+1,2}  - i\gamma_{j,2}\gamma_{j+2,1}\right)
	+ \frac{p-q}{4}\left( \gamma_{j+1,1}\gamma_{j+2,1} + \gamma_{j,2}\gamma_{j+1,2}\right) \right] \nonumber \\
    &+ \frac{1}{2}\left[(\alpha+\gamma)i\gamma_{1,1}\gamma_{1,2} + (\alpha-\gamma)\gamma_{1,1}\gamma_{2,1}\right]
    + \frac{1}{2}\left(\prod_{j=1}^{L+1} i\gamma_{j,1}\gamma_{j,2}\right)\left[(\delta+\beta)i\gamma_{L+1,1}\gamma_{L+1,2} - (\delta-\beta)\gamma_{L,2}\gamma_{L+1,2}\right].
\end{align}
\end{widetext}

The string of Majoranas $\prod_{j=1}^{L+1} \left(i\gamma_{j,1}\gamma_{j,2}\right) = (-1)^{L+1}\mathcal{P}_w$ equals, up to a sign, the parity operator $\mathcal{P}_w$ of Dirac fermions $w,w^\dagger$, which commutes with $H_0$. Thus, restricted to the sub-spaces of constant parity, the Hamiltonian $H_0$ becomes quadratic.

Note that $H_0$ in terms of the Majorana fermions $\gamma_{j,l}$ is acting non-trivially on the additional site $L+1$.  

To keep the algebra simpler let us consider from now on the case $p=1$ and $q=\gamma=\delta=0$. The following calculations can be straightforwardly generalized to arbitrary $p,q,\gamma,\delta$. Thus $H_0$ in terms of the Majorana fermions $\gamma$ simplifies to 
\begin{align}
	H_0 
    &= \sum_{j=1}^{L-1} \left(\sigma_{j+1}^+ \sigma_j^-  \right) + \alpha \sigma_1^+ + \beta \sigma_L^- \nonumber \\
    &= \frac{1}{2} \alpha [i\gamma_{1,1}\gamma_{1,2} + \gamma_{1,1}\gamma_{2,1}] \nonumber \\
    &+ \frac{1}{2}(-1)^{L+1}\mathcal{P}_w \beta [i\gamma_{L+1,1}\gamma_{L+1,2} + \gamma_{L,2}\gamma_{L+1,2}] \nonumber\\
	&+ \frac{1}{4}\sum_{j=1}^{L-1} \bigg[ (\gamma_{j,2},\ \gamma_{j+1,1}) \begin{pmatrix}
		1 &-i\\ i &1
	\end{pmatrix} 
	\begin{pmatrix}
		\gamma_{j+1,2}\\ \gamma_{j+2,1}
	\end{pmatrix} \bigg].
\end{align}

The eigenvalues of the $2\times2$-matrix are 0 and 2, while the eigenvectors are $(1, -i)^t$ and $(1, i)^t$, respectively. Thus the following pairing of Majorana fermions
\begin{equation}\label{eq:app_H0_obc_c_gamma}
	c_j^\dagger = \frac{1}{2}(\gamma_{j,2}-i\gamma_{j+1,1}), \quad
	c_j = \frac{1}{2}(\gamma_{j,2} + i\gamma_{j+1,1}),
\end{equation}
into Dirac fermions $c,c^\dagger$ drastically simplifies the bulk term. By identifying $\gamma_{L+2,1}=\gamma_{1,1}$ the pairing given by Eq.~\eqref{eq:app_H0_obc_c_gamma} turns the chain on sites 1 to $L+1$ into a ring, connecting site 1 and $L+1$. The Hamiltonian $H_0$ is given in terms of $c,c^\dagger$ as
\begin{align}
    H_0 = &\alpha(c_{L+1} - c_{L+1}^\dagger) c_1^\dagger 
	+ \sum_{j=1}^{L-1} \left[ c_j c_{j+1}^\dagger \right] \nonumber \\
	&+ (-1)^{L} \mathcal{P}_c \beta c_L(c_{L+1}+c_{L+1}^\dagger), \notag
\end{align}
where $\mathcal{P}_c$ denotes the parity of the Dirac fermions $c,c^\dagger$. This is Eq.\eqref{eq:H0_obc_c}.

\section{Diagonalizing $M_c$}\label{sec:app_Mc}
In this section we calculate the eigenvalues and eigenvectors of $M_c$ given by Eq.~\eqref{eq:H0_obc_Mc} and Eqs.(\ref{eq:H0_obc_A}-\ref{eq:H0_obc_C}) thereafter. We denote the eigenvalue equation by $M_c \bu = \lambda \bu$ with the $2L+2$ dimensional vector $\bu$. In terms of $\bu=(u_1,\dots,u_{L+1},u_1',\dots,u_{L+1}')$ the eigenvalue equation reads
\begin{align}
	\lambda u_1 &= -\alpha (u_{L+1} - u_{L+1}')   \label{eq:app_Mc_u1}    \\
	\lambda u_2 &= -u_1 					      \label{eq:app_Mc_u2}   \\
	&\dots 										   \notag \\
	\lambda u_{L-1} &= - u_{L-2}				   \label{eq:app_Mc_uLm1}\\
	\lambda u_{L} &= - u_{L-1} 					   \label{eq:app_Mc_uL}\\
	\lambda u_{L+1} &= -s\beta u_L - \alpha u_1'   \label{eq:app_Mc_uLp1}
\end{align}
and
\begin{align}
	\lambda u_1' &= u_2'                           \label{eq:app_Mc_up1}\\
	\lambda u_2' &= u_3'                           \label{eq:app_Mc_up2}\\
	&\dots 		\notag \\
	\lambda u_{L-1}' &= u_L'                       \label{eq:app_Mc_upLm1}\\
	\lambda u_L' &= \beta s(u_{L+1} +u_{L+1}')     \label{eq:app_Mc_upL}\\
	\lambda u_{L+1}'&=-\beta su_{L} + \alpha u_1'. \label{eq:app_Mc_upLp1}
\end{align}
Combining the Eqs.~\eqref{eq:app_Mc_u2}-\eqref{eq:app_Mc_uL} with $u_1,\dots,u_L$ and Eqs.~\eqref{eq:app_Mc_up1}-\eqref{eq:app_Mc_upLm1} with $u_1',\dots,u_L'$ recursively we get for $2\le j \le L$ 
\begin{equation}
	u_j = -\lambda^{-1} u_{j-1} = \dots = (-\lambda)^{-j+1} u_1
\end{equation}
and
\begin{equation}
	u_j' = \lambda u_{j-1}' =\dots =\lambda^{j-1} u_1'.
\end{equation}
Especially, the following holds
\begin{align}
	u_L &= (-\lambda)^{-L+1} u_1 \label{eq:app_Mc_uL_rec}  \\
	u_L' &= \lambda^{L-1} u_1'. \label{eq:app_Mc_upL_rec}
\end{align}
By substituting Eq.~\eqref{eq:app_Mc_uL_rec} and Eq.~\eqref{eq:app_Mc_upL_rec} into Eq.~\eqref{eq:app_Mc_uLp1} and Eq.~\eqref{eq:app_Mc_upLp1}, respectively, we get
\begin{align}
	u_1 &= \alpha \lambda^{-1} (-u_{L+1} + u'_{L+1}) \label{eq:app_Mc_u1_uLp1} \\
	u_{L+1} &=  \beta s(-\lambda)^{-L}u_1  -\alpha\lambda^{-1} u_1' \label{eq:app_mc_uLp1_u_1}\\
	u_1' &= \lambda^{-L} \beta s(u_{L+1} + u_{L+1}') \label{eq:app_Mc_up1_uLp1}  \\
	u_{L+1}' &= \beta s (-\lambda)^{-L} u_1 + \alpha \lambda^{-1} u_1'. \label{eq:app_mc_upLp1_u_1}
\end{align}
Adding and subtracting Eq.~\eqref{eq:app_mc_uLp1_u_1} and Eq.~\eqref{eq:app_mc_upLp1_u_1}, respectively, leads to
\begin{align}
	u_{L+1}+u_{L+1}' &= 2\beta s(-\lambda)^{-L} u_1 \\
	-u_{L+1}+u_{L+1}' &= 2\alpha\lambda^{-1}u_1',
\end{align}
which in turn implies that
\begin{align}
	u_1' &= 2 (-1)^L \lambda^{-2L}\beta^2 u_1 \\
	u_1 &= 2 \alpha^2 \lambda^{-2} u_1',
\end{align}
by using Eqs.~\eqref{eq:app_Mc_u1_uLp1} and ~\eqref{eq:app_Mc_up1_uLp1}. Combining the last two equations leads to
\begin{equation}
	u_1 = 4 (\alpha\beta)^2 (-1)^L \lambda^{-2L-2}u_1,
\end{equation}
which implies, for $u_1\neq 0$, the eigenvalue Eq.~\eqref{eq:H0_obc_roots}
\begin{equation}
	\lambda^{2(L+1)} = (-1)^L 4 (\alpha\beta)^2.
\end{equation}
The roots of this polynomial are given by
\begin{equation*}
	\lambda = (2\alpha\beta)^{\frac{1}{L+1}} \begin{cases}
		\exp\left(\frac{i\pi}{2L+2} 2k\right) 		&L \text{ even,} \\
		\exp\left(\frac{i\pi}{2L+2} (2k-1)\right) 	&L \text{ odd,}
	\end{cases}
\end{equation*}
where $k=1,\dots,2L+2$.

\begin{widetext}

\section{Bethe equations for ASEP with pbc}\label{sec:app_bethe}

In this section, we derive the Bethe equations presented in Sec.~\ref{sec:bethe_ansatz}.  These results extend the usual $U=1$ ASEP Bethe ansatz \cite{Gwa_Spohn_PRA1992_Bethe_ansatz} to the case of arbitrary $U$.

By $\ket{x_1,\dots,x_N}$ we denote the state of $N$ particles at position $x_1,\dots,x_N$. In the following, we let $x_1<\dots < x_N$ up to an overall shift in the indices. The wavefunction $\ket{\psi}$ in the basis of $\ket{x_1,\dots,x_N}$ is given by
\begin{equation}
	\ket{\psi} = \sum_{x_1<\dots<x_N} \psi(x_1,\dots,x_N) \ket{x_1,\dots,x_N},
\end{equation}
where $\psi(x_1,\dots,x_N)$ denotes the coefficient of $\ket{\psi}$ with respect to $\ket{x_1,\dots,x_N}$. Now, let $\ket{\psi}$ be an eigenstate of the generalized Markov matrix $H$ with eigenvalue $E$, i.e. $H\ket{\psi} = E \ket{\psi}$. Recall that we can write the generator matrix $H$ as
\begin{equation}\label{eq:appendix_M_gen_asep}
	H = \sum_{i=1}^L \left( p \sigma_i^- \sigma_{i+1}^+ + q \sigma_i^+ \sigma_{i+1}^- \right) + \frac{U}{4} \sum_{i=1}^L \left(\sigma_i^z \sigma_{i+1}^z - 1 \right).
\end{equation}

Let us first focus on the action of the off-diagonal term in eq.~\eqref{eq:appendix_M_gen_asep} on $\ket{x_1,\dots,x_N}$. It is easy to see that
\begin{equation}
	\sum_{i=1}^{L-1} \sigma_i^- \sigma_{i+1}^+ \ket{x_1,\dots,x_N}
	= \sum_{j=1}^{N-1} (1-\delta(x_{j+1}-x_j,1)) \ket{x_1,\dots,x_j+1,\dots,x_N}, 
 \end{equation}
and
\begin{equation}
 \sum_{i=1}^{L-1} \sigma_i^+ \sigma_{i+1}^- \ket{x_1,\dots,x_N}
	= \sum_{j=2}^{N} (1-\delta(x_{j}-x_{j-1},1)) \ket{x_1,\dots,x_j-1,\dots,x_N},
\end{equation}
where $\delta(x,y)$ equals one whenever $x=y$ and is zero otherwise. The remaining boundary terms are determined as follows. If $x_N\neq L$ then $\sigma_L^- \sigma_1^+ \ket{x_1,\dots,x_N}=0$, so let $x_N=L$. Then
\begin{align}
	\sigma_L^- \sigma_1^+ \ket{x_1,\dots,x_N}
	&= (1-\delta(x_1, 1)) \ket{1, x_1,\dots,x_{N-1}}\\
	&= (1-\delta(x_1-x_N \pmod L, 1)) \ket{x_1,\dots,x_{N-1}, X_N+1}
\end{align}
by identifying $\ket{x_1,\dots,x_{N-1}, L+1}=\ket{1,x_1,\dots,x_{N-1}}$. On the other hand, whenever $x_1\neq 1$ we have $\sigma_L^+\sigma_1^- \ket{x_1,\dots,x_N}=0$, while for $x_1=1$ we get
\begin{align}
	\sigma_L^+\sigma_1^- \ket{x_1,\dots,x_N} 
	&= (1-\delta(x_N,L)) \ket{x_2,\dots,x_N,L}\\
	&= (1-\delta(x_1 - x_N \pmod L, 1)) \ket{x_1-1,x_2,\dots,x_N},
\end{align}
where we identified $\ket{x_2,\dots,x_N,L} = \ket{0, x_2,\dots,x_N}$. Taking everything together we have
\begin{align}
	\sum_{i=1}^{L} \sigma_i^- \sigma_{i+1}^+ \ket{x_1,\dots,x_N}
	&= \sum_{j=1}^{N} (1-\delta(x_{j+1}-x_j \pmod L,1)) \ket{x_1,\dots,x_j+1,\dots,x_N}, \\
	\sum_{i=1}^{L} \sigma_i^+ \sigma_{i+1}^- \ket{x_1,\dots,x_N}
	&= \sum_{j=1}^{N} (1-\delta(x_{j}-x_{j-1} \pmod L,1)) \ket{x_1,\dots,x_j-1,\dots,x_N}.
\end{align}
The diagonal term in Eq.~\eqref{eq:appendix_M_gen_asep} acts on $\ket{x_1,\dots,x_N}$ as 
\begin{align}
	\sum_{i=1}^{L} \left( \sigma_i^z \sigma_{i+1}^z - 1\right) \ket{x_1,\dots,x_N}
	&= \sum_{j=1}^{N-1} \delta(x_{j+1} - x_j, 1) + \delta(x_1 - x_N, 1-L) - N \nonumber \\
	&= \sum_{j=1}^N \delta(x_{j+1}-x_j \pmod{L}, 1) - N,
\end{align}
where we note that $\sigma^z_i = 2n_i - 1$ and thus
\begin{align}
	\frac{1}{4}\sum_{i=1}^L \left(\sigma_i^z \sigma_{i+1}^z - 1\right)
	&= \frac{4}{4}\sum_{i=1}^L n_i n_{i+1} - \frac{2}{4} \sum_{i=1}^L n_i - \frac{2}{4} \sum_{i=1}^L n_{i+1} 
	=\left[\sum_{i=1}^L n_i n_{i+1} \right] - N.
\end{align}
Summarizing, the action of $H$ on $\ket{x_1,\dots,x_N}$ is
\begin{align}\label{eq:appendix_M_x}
	H\ket{x_1,\dots,x_N}
	&= p\sum_{j=1}^{N} (1-\delta(x_{j+1}-x_j \pmod L,1)) \ket{x_1,\dots,x_j+1,\dots,x_N} \nonumber\\
	&+ q\sum_{j=1}^{N} (1-\delta(x_{j}-x_{j-1} \pmod L,1)) \ket{x_1,\dots,x_j-1,\dots,x_N} \nonumber\\
	&- U\sum_{j=1}^N (1-\delta(x_{j+1}-x_j \pmod{L}, 1)) \ket{x_1,\dots,x_N}.
\end{align}

Now, consider the eigenvalue equation $H\ket{\psi} = E \ket{\psi}$,
\begin{align}
	H\ket{\psi} 
	&= \sum_{x_1<\dots<x_N} \psi(x_1,\dots,x_N) H\ket{x_1,\dots,x_N}
	= \sum_{x_1<\dots<x_N} \psi(x_1,\dots,x_N) E\ket{x_1,\dots,x_N}.
\end{align}
Let us concentrate on the term in Eq.~\eqref{eq:appendix_M_x} proportional to $p$
\begin{align}
	p\sum_{j=1}^{N} \sum_{x_1<\dots<x_N} \psi(x_1,\dots,x_N) (1-\delta(x_{j+1}-x_j \pmod L,1)) \ket{x_1,\dots,x_j+1,\dots,x_N}.
\end{align}
After a change of variables $\tilde{x}_i=x_i$ for $i\neq j$ and $\tilde{x}_j = x_j+1$ the above equation reads
\begin{align}
	p\sum_{j=1}^{N} \sum_{\tilde{x}_1<\dots<\tilde{x}_N} \psi(\tilde{x}_1,\dots,\tilde{x}_j-1,\dots,\tilde{x}_N) (1-\delta(\tilde{x}_{j}-\tilde{x}_{j-1} \pmod L,1)) \ket{\tilde{x}_1,\dots,\tilde{x}_N}.
\end{align}
Let us now focus on the term in Eq.~\eqref{eq:appendix_M_x} proportional to $q$. One finds with the change $\tilde{x}_j = x_j - 1 < x_{j+1} - 1 = \tilde{x}_{j+1}-1$, thus $\tilde{x}_{j+1}-\tilde{x}_j > 1$ and $\tilde{x}_{j-1} = x_{j-1} < x_j -1 = \tilde{x}_j$, that this term equals
\begin{align}
	q \sum_{\tilde{x}_1<\dots<\tilde{x}_N}\psi(\tilde{x}_1,\dots,\tilde{x}_j+1,\dots,\tilde{x}_N) (1-\delta(\tilde{x}_{j+1}-\tilde{x}_j \pmod L, 1)) \ket{\tilde{x}_1,\dots,\tilde{x}_N},
\end{align}
where the first constraint is realized via the delta term and the second constraint by the summation.

By orthogonality of $\ket{x_1,\dots,x_N}$ the eigenvalue equation $H\ket{\psi}=E\ket{\psi}$ turns into $\binom{L}{N}$ equations for the wavefunction coefficients
\begin{align}
	&p \sum_{j=1}^N \big(1-\delta(x_j-x_{j-1} \mod L, 1)) (\psi(x_1,\dots,x_j-1,\dots,x_N) - U\psi(x_1,\dots,x_N) \big) \nonumber \\
	+&q\sum_{j=1}^N \big(1-\delta(x_{j+1} - x_j \mod L, 1)) (\psi(x_1,\dots,x_j+1,\dots,x_N) - U\psi(x_1,\dots,x_N) \big) \nonumber\\
	&= E \psi(x_1,\dots,x_N). \label{eq:appendix_pq_ev}
\end{align}
Here we additionally used that $p+q=1$. Now, we make the ansatz for the wavefunction coefficient
\begin{equation}
	\psi(x_1,\dots,x_N) = \sum_{\tau\in S_N} A(\tau) \prod_{j=1}^N z_{\tau(j)}^{x_j},
\end{equation}
where the summation runs over all elements of the symmetric group $S_N$ and the $z_j$'s and $A(\tau)$'s are complex numbers. Let us consider a configuration $x_1<\dots<x_N$ where all particles have at least distance 1, i.e. no consecutive particles. Plugging the ansatz into the term proportional to $p$ results in
\begin{align}
	\sum_{j=1}^N\psi(x_1,\dots,x_j-1,\dots,x_N) 
	- U\psi(x_1,\dots,x_N) 
	&= \sum_{j=1}^N\sum_{\tau\in S_N} A(\tau) \left( z_{\tau(j)}^{x_j-1}\prod_{l=1;l\neq j}^N z_{\tau(l)}^{x_l} - U\prod_{l=1}^N z_{\tau(l)}^{x_l} \right) \\
	&= \sum_{\tau\in S_N} A(\tau) \prod_{l=1}^N z_{\tau(l)}^{x_l} \sum_{j=1}^N \left( z_{\tau(j)}^{-1} - U \right).
\end{align}
Similarly, one gets the analogous expression for the term proportional to $q$ with the change $z_{\tau(j)}^{-1}\to z_{\tau(j)}$. Thus Eq.~\eqref{eq:appendix_pq_ev} in terms of the Bethe ansatz reads
\begin{equation}
	E=\sum_{j=1} \left( pz_j+qz_j^{-1} - U \right),
\end{equation}
which is Eq.~\eqref{eq:bethe_energy} in Sec.~\ref{sec:bethe_ansatz}.
Now, consider a configuration $\ket{x_1,\dots,x_N}$ with two particles adjacent to each other. Then
\begin{equation}
	A(\dots,l,\dots,k,\dots) = - \frac{p+qz_lz_k-U z_l}{p+qz_lz_k-U z_k} A(\dots,k,\dots,l,\dots).
\end{equation}
The periodic boundary condition enforces $\psi(x_1,\dots,x_{N-1}, L+1)=\psi(1,x_1,\dots,x_{N-1})$, which implies
\begin{equation}
	A(\tau(1),\dots,\tau(N)) z_{\tau(N)}^L = A(\tau(N), \tau(1),\dots,\tau(N-1)).
\end{equation}
Combining both boundary constraints leads to the Bethe equations
\begin{equation}
	z_j^L = \prod_{k=1;k\neq j}^N \left(- \frac{p+qz_jz_k-U z_j}{p+qz_jz_k-U z_k} \right).
\end{equation}

In the case of TASEP with $q=1$ and $p=0$ the Bethe equations reduce to
\begin{align}
	z_j^L 
	&= \prod_{k=1;k\neq j}^N \left(- \frac{z_jz_k-U z_j}{z_jz_k-U z_k} \right) 
	= \frac{z_j^N}{(z_j-U)^N} (-1)^{N-1} \prod_{k=1}^N  \frac{z_k-U}{z_k},
\end{align}
so
\begin{equation}
	z_j^{L-N} (z_j-U)^N =(-1)^{N-1} \prod_{k=1}^N \frac{z_k-U}{z_k}.
\end{equation}
Denoting $Z_k = 2z_k -U$ we get
\begin{equation}
	(U + Z_j)^{L-N} (U-Z_j)^N = -2^L \prod_{k=1}^N \frac{Z_k-U}{Z_k+U}.
\end{equation}

\end{widetext}

\section{Solving the Bethe equations numerically}\label{sec:app_bethe_numerical}
\begin{figure}
	\begin{center}
		\includegraphics[width=\columnwidth]{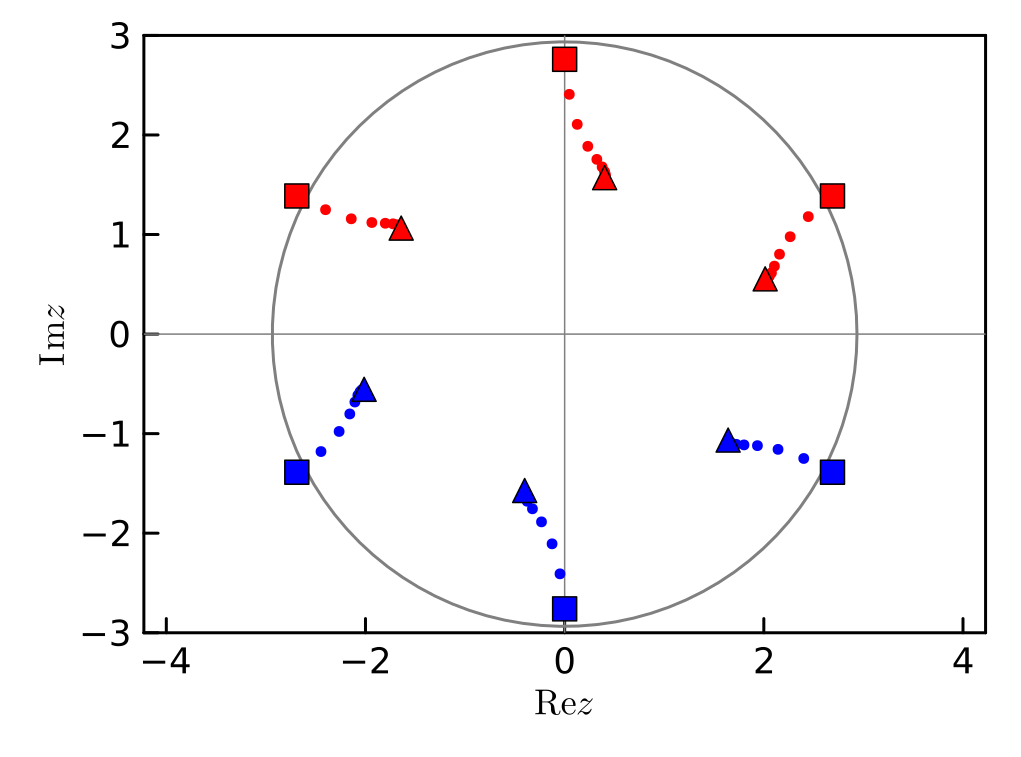}
		\caption{Visualization of solving the Bethe Eqs.~\eqref{eq:bethe_tasep_recurrent} of TASEP (pbc) for $L=6$ and $N=3$. All markers are roots of the polynomial $P$ (Eq.~\eqref{eq:bethe_polynomial}) for different $Y$. The outer (square) markers are the roots for initial $Y^{(1)}=10\times 2^L$, the inner (triangles) markers for $Y$ converged, and the circles denote roots of $P$ for intermediate $Y$. Red markers (upper complex plane) are chosen to calculate the next $Y$. Gray circle has radius $|Y^{(1)}|^{1/L}$. \label{fig:solve_bethe_roots}}
	\end{center}
\end{figure}

In this section, we will describe how to self-consistently solve the Bethe equations numerically. We will mostly follow the approach in \cite{Golinelli_Mallick_JourPhysA2006ASEP} with some additional tweaks.

Restricting to $p=1$ and $q=0$ reduces the difficulty of solving the Bethe equations considerably because the right-hand side of Eq.~\eqref{eq:bethe_tasep_recurrent} does not depend on $j$, as does the right-hand side of Eq.~\eqref{eq:bethe_recurrent} for general $p,q$. 

Consider the polynomial $P(z)$,
\begin{equation}\label{eq:bethe_polynomial}
	P(z) = \left(U+z\right)^{L-N}\left(U-z\right)^N - Y,
\end{equation}
where $Y$ denotes an arbitrary complex number and let us denote the right-hand side of Eq.~\eqref{eq:bethe_tasep_recurrent} by
\begin{equation}
    \tilde{Y}(Z_1,\dots,Z_N) = 2^L \prod_{k=1}^N \frac{Z_k-U}{Z_k+U}.
\end{equation}
Then every solution $Z_1,\dots,Z_N$ of Eq.~\eqref{eq:bethe_tasep_recurrent} are roots of the polynomial $P$ with $Y=\tilde{Y}(Z_1,\dots,Z_N)$. To find a solution to the Bethe equations one first calculates the roots $Z_1^{(1)},\dots,Z_L^{(1)}$ of $P$ for an initial $Y^{(1)}$. Of these $L$ roots of $P$ one chooses $N$ roots, $Z_1^{(1)},\dots,Z_N^{(1)}$, and evaluates the next $Y^{(2)} = \tilde{Y}(Z_1^{(1)},\dots,Z_N^{(1)})$. Again, the roots $Z_1^{(2)},\dots,Z_L^{(2)}$ of $P$ with $Y=Y^{(2)}$ are calculated and $N$ roots $Z_1^{(1)},\dots,Z_N^{(1)}$ are chosen to evaluate the next $Y^{(3)}=\tilde{Y}(Z_1^{(2)},\dots,Z_N^{(2)})$. This procedure is then iterated until convergence all of the $N$ chosen roots is reached, $Z_j^{(l)} \approx Z_j^{(l+1)}$ for all $1\le j\le N$. 

The convergence of this procedure presupposes consistency of the choice of the $N$ roots out of $L$ roots of the polynomial $P$ \cite{Motegi_etal_PRE_2012_bethe_dynamics, Prolhac_JournStatMech2015_fluct_dev_TASEP}. The first choice of $Z_1^{(1)},\dots,Z_N^{(1)}$ out of $Z_1^{(1)},\dots,Z_L^{(1)}$ is arbitrary. Subsequent roots $Z_1^{(l)},\dots,Z_N^{(l)}$ are chosen to be closest to the previous roots
\begin{equation}
    Z_j^{(1)} = \operatorname*{argmin}_{Z_k^{(l)}:1\le k \le L} |Z_k^{(l)} - Z_j^{(l-1)}|,
\end{equation}
where the minimum runs over all roots $Z_1^{(l)},\dots,Z_L^{(l)}$ of $P$ with $Y=Y^{(l)}$. If multiple $Z_k^{(l)}$ are close to $Z_j^{(l-1)}$ we do not update $Y^{(l+1)}$ with $Z_j^{(l)}$ but with a linear combination of $Z_j^{(l)}$ and $Z_j^{(l-1)}$, i.e. $Y^{(l+1)} = \tilde{Y}(\dots, dY\ Z_j^{(l)} + (1-dY) Z_j^{(l-1)}, \dots)$ where $0<dY\le 1$ denotes the fraction of interpolation between $Z_j^{(l)}$ and $Z_j^{(l-1)}$.

The above-described procedure typically leads to convergence of $Z_1^{(l)},\dots,Z_N^{(l)}$ and thus to a solution of the Bethe Eqs.~\eqref{eq:bethe_tasep_recurrent}. In Fig.~\ref{fig:solve_bethe_roots} we show the roots $Z_1^{(l)},\dots,Z_6^{(l)}$ obtained during the above algorithm for $L=6$ and $N=3$. The square markers denote the initial $Z_1^{(1)},\dots,Z_6^{(1)}$ with $Y^{(1)}=10\times 2^L$, while the triangles denote the final and converged $Z_1^{(end)},\dots,Z_6^{(end)}$ (relative or absolute error of Eq.~\eqref{eq:bethe_tasep_recurrent} $<10^{-3}$). The circles indicate intermediate roots. Initially, the 3 red squares (upper half-plane) are chosen as $Z_1^{(1)},\dots,Z_3^{(1)}$, and subsequent roots (upper half-plane in red) according to their previous closest roots. For visualization purposes, $dY$ was chosen to be $dY=0.5$.

To find all solutions to the Bethe Eqs.~\eqref{eq:bethe_tasep_recurrent} systematically we use different combinations of initial $Y^{(1)}$ and initial root choices. Namely, we typically choose $Y^{(1)}$ with $|Y^{(1)}|^{1/L} \gg U$. This ensures that the roots of $P$ with $Y=Y^{(1)}$ are close to the circle with radius $|Y^{(1)}|^{1/L}$. In Fig.~\ref{fig:solve_bethe_roots} the roots of $P$ for $Y=Y^{(1)}=10\times 2^6$ denoted by the square markers are close to the circle with radius $2\times 10^{1/6}\approx 2.9$. Then we solve the Bethe equations for every combination of $N$ roots out of $L$. This typically gives us almost all solutions of the Bethe Eqs.~\eqref{eq:bethe_tasep_recurrent}. By iterating this procedure for a handful of initial $Y^{(1)}$ we found all Bethe roots for the systems we investigated (up to $L=22$).

\bibliography{lit_asep}

\end{document}